\documentclass[sigconf, preprint, anonymous=false]{acmart}

\makeatletter
\def\@copyrightspace{\relax}
\makeatother

\renewcommand\footnotetextcopyrightpermission[1]{} %
\usepackage{booktabs} %
\usepackage{csquotes}
\usepackage{longtable}
\usepackage{multirow}

\usepackage{listings}
\colorlet{punct}{red!60!black}
\definecolor{delim}{RGB}{20,105,176}
\colorlet{numb}{magenta!60!black}
\colorlet{more}{green}

\definecolor{keyword}{HTML}{2771a3}
\definecolor{pattern}{HTML}{b53c2f}
\definecolor{string}{HTML}{be681c}
\definecolor{relation}{HTML}{7e4894}
\definecolor{variable}{HTML}{107762}
\definecolor{comment}{HTML}{8d9094}

\lstdefinelanguage{json}{
	basicstyle=\normalfont\ttfamily,
	showstringspaces=false,
	breaklines=true,
	numbers=left,
	numberstyle=\scriptsize,
	stepnumber=5,
	numbersep=8pt,
	literate=
	*{0}{{{\color{numb}0}}}{1}
	{1}{{{\color{numb}1}}}{1}
	{2}{{{\color{numb}2}}}{1}
	{3}{{{\color{numb}3}}}{1}
	{4}{{{\color{numb}4}}}{1}
	{5}{{{\color{numb}5}}}{1}
	{6}{{{\color{numb}6}}}{1}
	{7}{{{\color{numb}7}}}{1}
	{8}{{{\color{numb}8}}}{1}
	{9}{{{\color{numb}9}}}{1}
	{<}{{{\color{more}<}}}{1} %
	{>}{{{\color{more}>}}}{1} %
	{:}{{{\color{punct}{:}}}}{1}
	{,}{{{\color{punct}{,}}}}{1}
	{\{}{{{\color{delim}{\{}}}}{1}
	{\}}{{{\color{delim}{\}}}}}{1}
	{[}{{{\color{delim}{[}}}}{1}
	{]}{{{\color{delim}{]}}}}{1},
}

\lstdefinestyle{ebnf}{language=sh,
	morekeywords={super},
		numbers=left,
	numberstyle=\scriptsize,
	stepnumber=5,
	numbersep=8pt,
	literate=
	*{0}{{{\color{numb}0}}}{1}
	{1}{{{\color{numb}1}}}{1}
	{2}{{{\color{numb}2}}}{1}
	{3}{{{\color{numb}3}}}{1}
	{4}{{{\color{numb}4}}}{1}
	{5}{{{\color{numb}5}}}{1}
	{6}{{{\color{numb}6}}}{1}
	{7}{{{\color{numb}7}}}{1}
	{8}{{{\color{numb}8}}}{1}
	{9}{{{\color{numb}9}}}{1}
	{:}{{{\color{punct}{:}}}}{1}
	{=}{{{\color{punct}{:=}}}}{1} %
	{;}{{{\color{punct}{;}}}}{1}
	{,}{{{\color{punct}{,}}}}{1}
	{\{}{{{\color{delim}{\{}}}}{1}
	{\}}{{{\color{delim}{\}}}}}{1}
	{[}{{{\color{delim}{[}}}}{1}
	{]}{{{\color{delim}{]}}}}{1},
}

\usepackage[]{todonotes}

\acmConference[ACM FAccT 2021 (Accepted)]{}{}{}
\acmYear{2021}

\newcommand{\req}{\textcolor{red}{*}}
\usepackage{array}
\newcolumntype{H}{>{\setbox0=\hbox\bgroup}c<{\egroup}@{}}

\usepackage{multicol}
\usepackage{csquotes}
\usepackage{footnote}

\usepackage{hyperref}
\setcopyright{none} 

\usepackage[false]{anonymous-acm}

\begin{document}
\title{TILT: A GDPR-Aligned %
Transparency Information Language and Toolkit for Practical Privacy Engineering}

\author{Elias Grünewald}
\orcid{0000-0001-9076-9240}
\affiliation{%
  \institution{Technische Universität Berlin\\ Information Systems Engineering Research Group}
  \city{Berlin} 
  \state{Germany} 
}
\email{gruenewald@tu-berlin.de}

\author{Frank Pallas}
\orcid{0000-0002-5543-0265}
\affiliation{%
  \institution{Technische Universität Berlin\\ Information Systems Engineering Research Group}
  \city{Berlin} 
  \state{Germany} 
}
\email{frank.pallas@tu-berlin.de}

\begin{abstract}
In this paper, we present TILT, a transparency information language and toolkit explicitly designed to represent and process transparency information in line with the requirements of the GDPR and allowing for a more automated and adaptive use of such information than established, legalese data protection policies do. 

We provide a detailed analysis of transparency obligations from the GDPR to identify the expressiveness required for a formal transparency language %
intended to meet respective legal requirements. In addition, we identify a set of further, non-functional requirements that need to be met to foster practical adoption in real-world (web) information systems engineering. On this basis, we specify our formal language and present a respective, fully implemented  toolkit around it. We then evaluate the practical applicability of our language and toolkit and demonstrate the additional prospects it unlocks through %
two different use cases: %
\textit{a)} the inter-organizational analysis of personal data-related practices allowing, for instance, to uncover data sharing networks based on explicitly announced transparency information and \textit{b)} the presentation of formally represented transparency information to users through novel, more comprehensible, and potentially adaptive user interfaces, heightening data subjects' actual informedness about data-related practices and, thus, their sovereignty. 

Altogether, our transparency information language and toolkit allow -- differently from previous work -- to express transparency information in line with actual legal requirements and practices of modern (web) information systems engineering and thereby pave the way for a multitude of novel possibilities to heighten transparency and user sovereignty in practice.
\end{abstract}

\begin{CCSXML}
<ccs2012>
   <concept>
       <concept_id>10010405.10010455.10010458</concept_id>
       <concept_desc>Applied computing~Law</concept_desc>
       <concept_significance>500</concept_significance>
       </concept>
   <concept>
       <concept_id>10002951.10003227</concept_id>
       <concept_desc>Information systems~Information systems applications</concept_desc>
       <concept_significance>500</concept_significance>
       </concept>
   <concept>
       <concept_id>10002951.10003260.10003309</concept_id>
       <concept_desc>Information systems~Web data description languages</concept_desc>
       <concept_significance>500</concept_significance>
       </concept>
   <concept>
       <concept_id>10011007.10011006.10011039</concept_id>
       <concept_desc>Software and its engineering~Formal language definitions</concept_desc>
       <concept_significance>500</concept_significance>
       </concept>
   <concept>
       <concept_id>10011007.10011006.10011050</concept_id>
       <concept_desc>Software and its engineering~Context specific languages</concept_desc>
       <concept_significance>500</concept_significance>
       </concept>
   <concept>
       <concept_id>10002978.10003029.10011150</concept_id>
       <concept_desc>Security and privacy~Privacy protections</concept_desc>
       <concept_significance>500</concept_significance>
       </concept>
 </ccs2012>
\end{CCSXML}

\ccsdesc[500]{Applied computing~Law}
\ccsdesc[500]{Information systems~Information systems applications}
\ccsdesc[500]{Information systems~Web data description languages}
\ccsdesc[500]{Software and its engineering~Formal language definitions}
\ccsdesc[500]{Software and its engineering~Context specific languages}
\ccsdesc[500]{Security and privacy~Privacy protections}

\keywords{Data transparency, GDPR, data protection, privacy by design, legal informatics, privacy engineering}

\maketitle

\sloppy

\section{Introduction}%

Transparency has been a core principle in philosophical, legal, and technical deliberations around privacy\footnote{Being well aware of the slightly different notions between ``Privacy'' and ``Data Protection'', we use these terms interchangeably herein.} for decades. This particularly applies to data privacy in everyday digital life: To be able to act in a sovereign and self-determined way and actually make \emph{informed} choices, individuals need to have sufficient knowledge about the actual facts and givens regarding the processing of personal data, including, e.g., which party collects what personal data for which purposes, how long this data is (going to be) stored, et cetera. Data protection laws and regulations around the world -- such as the European General Data Protection Regulation (GDPR) \cite{gdpr} or the California Consumer Privacy Act (CCPA) \cite{ccpa} -- therefore explicitly include transparency rules for data processing parties, obligating them to reveal respective information to the data subjects. 

These obligations are today typically fulfilled through respective transparency parts of written privacy policies. Such privacy policies do, however, exhibit several shortcomings that severely limit their actual reception and comprehension on the side of data subjects: First of all, privacy policies are often long, complex, and written in legalese language, making it hard for data subjects to locate transparency-related information and actually understand them correctly \cite{ReidenbergDisagreeablePrivacyPolicies2015,Linden2018}. Second, different privacy policies employ different logical structures and vocabularies for factually similar statements, causing significant reading and decoding efforts for every new policy to be understood \cite{donald-cranor2008cost-reading}. Such practice structurally discriminates people who are less privacy-literate \cite{trepte2015}. Finally, data transfers towards other parties are now a broadly established practice. Data subjects wanting to see ``the whole picture'' in such contexts therefore have to read and understand multiple policies and establish the logical interdependencies between them themselves. 

Together with the ever-increasing number of services used (and, thus, of privacy policies to be read), this leads to a state where privacy policies, including respective transparency statements, are not read anymore before using a particular service and/or consenting to a certain collection and use of personal data \cite{obar2018}. Under such conditions, transparency statements increasingly degenerate into rather self-serving formal compliance exercises instead of actually fostering data subjects' informedness of decisions and self-sovereign conduct with regard to privacy.

For other areas of privacy, dedicated technologies have repeatedly been demonstrated to be capable of lowering the cognitive and administrative effort required on the side of data subjects and thereby of re-aligning the actual real-world practice to the (still valid) original intentions behind privacy regulations. Technologies for digitally mediated and partially automated consent provision are one example: They lower individuals' need to care about every single act of data sharing while still allowing them to fine-tune their consent to individual preferences in line with legal requirements for specificity, thus counteracting ongoing trends towards overly broad consent provision \cite{UlbrichtYaPPLLightweightPrivacy2018}.

Following the paradigm of privacy by design, which basically requires \emph{all} privacy principles to be appropriately reflected in technology\footnote{See, for instance, Art. 25 GDPR requiring technical measures ``designed to implement data protection principles'' and, thus, covering all principles mentioned in Art. 5.}, we aim to achieve the same for the principle of transparency herein. Our goal is thus to provide technical artifacts that significantly lower user-side efforts required for gathering and comprehending transparency information while at the same time providing the capabilities necessary to meet regulatory requirements and, thus, to actually be applicable in practice. An indispensable precondition for such artifacts is the capability to represent relevant transparency information in a structured, machine-readable format and the possibility to easily employ this representation in real-world information systems. %
For this purpose, we particularly provide the following contributions:

\begin{itemize}
    \item An in-depth analysis of transparency information that needs to be expressible to meet the requirements of the GDPR
    \item A formal specification of a structured, machine-readable language meeting these expressiveness requirements -- the ``Transparency Information Language''
    \item Two fully implemented libraries for widely used programming languages (Python and Java) allowing to easily process and manage respective representations in real-world information systems -- the most important parts of the ``Toolkit'' around our language
    \item A demonstration of the opportunities arising from the machine-readable representation and technical mediation of transparency information through two exemplary case-studies for \textit{a)} inter-organizational analyses of stated data transfers and \textit{b)} novel, adaptive user interfaces
\end{itemize}

We thus explicitly follow an engineering-driven approach, providing novel, practically usable technological artifacts that address a currently open socio-technical challenge. Our respective considerations and contributions unfold as follows: Section~\ref{sec:related-work} summarizes related work on transparency enhancing technologies and identifies requirements for the aspired language and toolkit. Afterwards, section~\ref{sec:expressiveness} distills the necessary expressiveness of a transparency information language based on structured GDPR study. Based on these deliberations, the new transparency information language is formally defined and technically implemented (see section~\ref{sec:language design and implementation}). In section~\ref{sec:toolkit}, the language is practically embedded into a toolkit with storage and interoperability functionalities. In section~\ref{sec:exemplary}, we then demonstrate two exemplary applications, which enhance transparency using the language and toolkit components. Eventually %
possible future work is conceptualized in section \ref{sec:discussion}. Finally, section~\ref{sec:conclusion} concludes.

\section{Related Work \& Requirements}
\label{sec:related-work}

Technical approaches to (privacy-related) transparency are commonly discussed under the term of transparency-enhancing technologies (TETs). Such TETs exist in a broad variety of forms addressing transparency from many different angles. In particular, TETs comprise such diverse technologies as generating a location history based on social media entries \cite{creepy}, visualizing data exports \cite{Karegar2016}, phone sensor permission management \cite{semadroid}, or in-browser cookie tracking \cite{lightbeam}. For categorizing these TETs, different taxonomies exist \cite{hedbom2009, janic2013}. Zimmerman \cite{zimmermann2015} particularly distinguishes \emph{ex-ante}, \emph{realtime}, and \emph{ex-post} TETs, with ex-ante TETs providing information about a processor's \emph{intended} data collection and processing before the collection. 
Our aspired goal of making transparency-related statements more accessible and comprehensible for data subjects to heighten the informedness of their decisions clearly falls into this ex-ante category. Often mentioned ``privacy dashboards'' providing an integrated view to current and past possession and processing of data %
\cite{bier2016, angulo2015, Raschke2018}, in contrast, would fall into the ex-post category and thus clearly differ from our intent. Within the subdomain of ex-ante TETs, in turn, the herein addressed problems associated with written transparency statements %
not meeting their originally intended goals are particularly subject to two research strands: Automated knowledge extraction from privacy policies and pre-existing transparency languages. Both shall be introduced briefly before distilling requirements for our aspired language and toolkit.

\subsection{Privacy Policy Knowledge Extraction}

One established approach for %
technically addressing transparency-related problems is to extract %
respective information from pre-existing policies by means of natural language processing (NLP) and to then provide additional functionalities on top of the so-extracted data. This particularly includes attempts based on extensive static rules and named entity recognition  \cite{costante2013extraction} as well as more dynamic ones employing methods of crowdsourcing and machine-learning \cite{polisis}. So far, however, these typically focus on merely identifying blocks from a privacy policy dedicated to different subjects (like ``collection'') instead of actually \emph{extracting} information at a sufficient level of detail \cite{u3p, chang-policy-extraction} or achieve accuracies that are insufficient to be considered a reasonable basis for providing functionality with legal relevance on top of them \cite{polisis, costante2013extraction}. In addition, respective endeavors do typically not pay explicit regard to the actual and detailed regulatory requirements. A structured, machine-readable and interoperable representation that can be used for a multitude of different purposes is typically also out of scope.

The higher-level functionality aspired and sometimes even implemented in above-mentioned projects may, however, serve as blueprint for what we intend to facilitate with our structured transparency language and toolkit. In particular, this includes enhanced graphical presentations intended to improve comprehensibility \cite{polisis}, chatbots providing answers about a company's personal data practices \cite{pribots}, or automatically generated icon representations allowing for a quick and intuitive overview of personal data practices \cite{fukushima2018icon-representation}.

\subsection{Transparency Languages}

Given the insufficient prospects of starting with pre-existing, textual transparency information from privacy policies, an alternative approach is to start with a structured and machine-readable representation of the transparency information to be provided. %
Frameworks proposed in this regard so far -- like respective parts of the W3Cs early P3P standard \cite{cranor2003} or industry-driven initiatives like IAB's Transparency \& Consent Framework \cite{iab-tcf} -- do, however
lack the expressiveness actually required by privacy regulation. For instance, P3P provided a severely limited vocabulary of pre-defined purposes which does not suffice to express actually occuring purposes in the required level of specificity \cite[see, e.g.,][]{UlbrichtYaPPLLightweightPrivacy2018} while the IAB Europe Framework completely neglects several information obligations (e.g. rights to access, automated decision making etc.) and claims registration fees for participation. More recent policy languages \cite[e.g.,][]{gerl2019,special-usage-language} come closer to the legally required expressiveness aspired herein. However, these typically try to cover \emph{all} aspects of a privacy policy from the outset, resulting in a considerable level of complexity (e.g. considering the problem of de-identification mixed with transparency obligations) and/or still resemble the concept of limited, fixed vocabularies known from earlier proposals. At the same time, they typically lack publicly available and easily adoptable (reference) implementations in the form of re-usable libraries, severely limiting their practical applicability in real-world (web) information systems. The effective adoption of privacy languages also has been detained because of missing content validation and (not even rudimentary) user interfaces.

Altogether, none of the %
transparency-related technologies and languages proposed so far thus satisfies our intent to represent transparency information as required by %
the GDPR ex-ante in a machine-readable form and in line with actual legal requirements while at the same time being easily usable in practice.

\subsection{Requirements}

Based on the above and on other works in the domain of information systems engineering in general and practice-oriented privacy engineering in particular \cite[esp.][]{pbac2020}, we can identify a set of requirements our aspired language and toolkit have to fulfill. These comprise specifications regarding \emph{what} functionality is to be implemented (functional requirements, FR) as well as rather non-functional requirements (NFR) describing additional characteristics that the language and toolkit must provide in order to ease and 
foster practical adoption.

\paragraph{Req. 1 (FR): Sufficient expressiveness to meet legal obligations} The most important requirement is a functional one. Our language must be capable of expressing all transparency information required by the GDPR in a sufficiently specific form. Only when this is the case, our language can actually form a sustainable basis for novel technical approaches to transparency that still unfold legal relevance and, thus, overcome the limitations of previous approaches like P3P (see above). The %
details constituting this requirement will be elaborated further in section \ref{sec:expressiveness}.

\paragraph{Req. 2 (FR): Possible standardization of commonly used transparency information items}  Another goal regards the standardization of as many transparency information items as possible. Through the introduction of structured-types, nomenclature and format conventions, the content becomes more accessible and interoperable. Besides fostering the independent development of different components on top of commonly shared semantics, this would also allow to create, for instance, templates for transparency declarations, standardized mappings to textual representations, etc.

\paragraph{Req. 3 (FR): Support for inter-system communication and versioned persistence} Our language is explicitly intended to be used in online contexts, with transparency information being exchanged between multiple sub-components of an overall system, external services, user-side applications, etc. Capabilities for communicating and storing respective representations as well as for referring to them across component- and system boundaries in a standardized manner are therefore indispensable. In particular, this should be possible in a version-aware form, allowing to update transparency information while still keeping previous versions available.

\paragraph{Req. 4 (FR): Application- or Domain-Specific Extensibility} The benefits of standardization laid out in req. 2 notwithstanding, specific application domains, service providers or even single applications might raise additional transparency requirements that cannot reasonably be covered in a generalized manner. This could, for instance, be the case because of sector-specific regulatory obligations going beyond those from the GDPR. To also facilitate the usage of our artifacts in such settings, they must allow to implement respective extensions on top of their underlying, general functionality. In particular, this should be possible in a way that preserves the applicability of extensions in case of our general components being updated (thus speaking against domain-specific forks, for example).

\paragraph{Req. 5 (NFR): Implementation as reusable artifact} A core principle of software engineering is to not implement similar functionality multiple times but to encapsulate respective functionality in re-usable components as far as possible. Besides avoiding inconsistencies between different implementations and allowing to focus development and maintenance efforts on just one target, this also fosters wider adoption and simpler integration into real-world systems. Our technical artifacts should therefore be provided in the form of reusable artifacts that can be integrated into relevant development environments as seamlessly as possible.

\paragraph{Req. 6 (NFR): Developer-Friendliness and and low implementation overhead} Besides the provision as reusable artifacts, developer-friendliness and low implementation efforts are also crucial for facilitating adoption in practice. On the one hand, this is the case because developers will try to avoid sophisticated and hard-to use technical mechanisms while intuitive and easily integratable ones might be adopted out of developers' intrinsic motivation or mere curiosity \cite{pbac2020}. On the other hand, developer-friendliness and ease of implementation are also relevant for a more formal reason: Art. 25 of the GDPR requires ''appropriate technical [...] measures'' to be implemented for materializing data protection principles (including transparency). The appropriateness, in turn, depends (amongst other factors) on ``the costs of implementation''. The lower the implementation efforts (and, thus, costs) for adopting a novel technical mechanism are, the more likely is this adoption therefore to be considered obligatory. Developer-friendliness and low implementation efforts thus also have legal implications.

\section{Expressiveness Requirements} \label{sec:expressiveness}

To unfold legal relevance as a possible replacement for written transparency statements and, thus, to provide actual societal and business value, a transparency language must be capable of expressing all legally required transparency information at a sufficient level of detail. %
We refer to this capability under the term of \emph{expressiveness}. Given its prominent role as a blueprint for privacy and data protection regulations worldwide, we extract respective  requirements from the European General Data Protection Regulation (GDPR) \cite{gdpr}.

Here, transparency requirements are set out in Articles 12-15. More precisely, Art. 12 requires controllers to provide transparency information to data subjects ``in a concise, transparent, intelligible and easily accessible form''. Even though implicitly assuming the provision %
through traditional, purely textual policies, the GDPR also foresees alternative communicative channels such as standardized and machine-readable icons (see Art. 12 (7)).

The transparency information to be provided to data subjects proactively, in turn, are laid out in Art. 13 and 14 in more detail for different settings (distinguishing cases where personal data is collected from the data subject from those where it is obtained from elsewhere). Beyond these, Art. 15 defines information to be provided to data subjects upon request, particularly also including the personal data itself. Leaving aside the latter (which is overly complex and diverse across different data controllers to be covered by a uniform language as aspired herein), the transparency obligations from Art. 13-15 strongly overlap and shall serve as the basis for determining the required expressiveness herein. 

Besides the mentioned ones, Art. 30 -- obligating data controllers to maintain ``records of processing activities'' -- also exhibits strong overlaps regarding the information to be kept in such records. Even though serving the principle of \emph{accountability} and not the one of \emph{transparency} (cf. Art. 5(2)), our aspired language may thus also play a role in this regard later. We therefore include the requirements from Art. 30 in our analysis. The resulting summary of all relevant articles is provided in table \ref{tab:summary_transp_information}.

\begin{table}[t!]
  \caption{Transparency information obligations by the GDPR}
  \label{tab:summary_transp_information}

\small 
    \setlength{\tabcolsep}{4pt}    
    \noindent \begin{tabular}[]{p{0.8cm}p{0.8cm}p{0.8cm}p{0.8cm}p{3.5cm}H}
    \textbf{Reference(s)} & & & & \textbf{Transparency information} & \textbf{Description}\\
    \midrule
    13 (1a) & 14 (1a) & & 30 (1a) & Controller & Name and contact details
    of the person responsible and, if applicable, of the
    representative\tabularnewline	\midrule
    13 (1b) & 14 (1b) & & 30 (1a) & Data protection officer & Contact
    details of the data protection officer\tabularnewline	\midrule
    13 (1c) & 14 (1c) & 15 (1a) & 30 (1b) & Purposes & Purposes for which
    the personal data are to be processed\tabularnewline	\midrule
    13 (1c) & 14 (1c) & & & Legal basis & Legal basis for the
    processing\tabularnewline	\midrule
    13 (1d) & 14 (2b) & & & Legitimate interests & Legitimate interests pursued by the controller or a third party when processing in accordance with Article 6 (1f)\tabularnewline	\midrule
    13 (1e) & 14 (1e) & 15 (1c) & 30 (1d) & Recipient (categories) &
    Recipients or categories of recipients of the personal
    data, if any\tabularnewline	\midrule
    13 (1f) & 14 (1f) & 15 (1c) & 30 (1e) & Third country transfer & The
    controller's intention to transfer the personal data to a third country
    or an international organization\tabularnewline	\midrule
    13 (1f) & 14 (1f) & 15 (2) & 30 (1e) & Adequacy (third country) &
    Existence or absence of an adequacy decision by the Commission\tabularnewline	\midrule
    13 (1f) & 14 (1f) & 15 (2) & 30 (1e) & Access and Data portability & Reference to the appropriate or
    reasonable guarantees and the possibility of obtaining a copy of the
    personal data or where it is available in the case of transfers under
    Article 46 or Article 47 or Article 49 (1) second
    subparagraph\tabularnewline	\midrule
    13 (2a) & 14 (2a) & 15 (1d) & 30 (1f) & Retention or storage
    criteria & Duration for which the personal data is stored or, if this is
    not possible, the criteria for determining this duration\tabularnewline	\midrule
    13 (2b) & 14 (2c) & 15 (1e) & & Right to request access & There is a right to information
    from the person responsible about the personal data
    concerned\tabularnewline	\midrule
    13 (2b) & 14 (2c) & 15 (1e) &  & Right to correction or deletion &
    There is a right to correction or deletion or restriction of processing
    or a right to object to processing\tabularnewline	\midrule
    13 (2b) & 14 (2c) & 15 (1e) &  & Right to data portability & There
    is a right to data portability\tabularnewline	\midrule
    13 (2c) & 14 (2d) & & & Right to withdraw consent & There is a
    right to withdraw consent at any time without affecting the lawfulness
    of processing based on consent before the withdrawal\tabularnewline	\midrule
    13 (2d) & 14 (2e) & 15 (1f) & & Right to complaint & Right to lodge a
    complaint with a supervisory authority\tabularnewline	\midrule
    13 (2e) & & & & Necessity/Non-disclosure & Information on whether the provision of personal data
    is required by law or contract or is required for the conclusion of a
    contract, whether the data subject is obliged to provide the personal
    data and what possible consequences the non-provision would
    have\tabularnewline	\midrule
    13 (2f) & 14 (2g) & 15 (1h) &  & Automated decision making &
    The existence of automated decision-making, including profiling, and
    meaningful information about the logic involved, as well as the scope
    and intended effects of such processing for the data
    subject\tabularnewline	\midrule
    & 14 (2f) & & & Sources & Source from which the personal data originate
    and, if applicable, whether it originates from publicly accessible
    sources\tabularnewline	\midrule
    13 (3) & & & & Notification on purpose change & If the person responsible intends to process the personal
    data for a purpose other than that for which the personal data was
    collected, he shall provide the data subject with information about this
    other purpose and all other relevant information in accordance with
    paragraph 2 before this processing\tabularnewline	\midrule
    & & & 30 (1c) & Data subjects/Data disclosed & Description of
    the categories of data subjects and the categories of personal
    data
\end{tabular}
\end{table}

Each of these transparency requirements can be translated into a corresponding building block for our aspired transparency information language. The required data fields and respective data types to be included in these blocks result either from the GDPR itself or from technical considerations regarding unambiguousness. In some cases, the exact data fields and types to be provided are not externally given and we had to make reasonable determinations to allow for a well-specified formal representation in subsequent steps. Due to space constraints, the considerations behind each and every field and type cannot be laid out herein. Two building blocks shall, however, exemplarily be examined in some more detail:

\paragraph{Data Protection Officer:} Art. 13 and 14 of the GDPR require ``the contact details of the data protection officer'' to be provided whereas Art. 30 requires ``the \emph{name and} contact details'' (emph. added) to be included in the record of processing activities. As no further definition is provided regarding \emph{which} contact details are required, we had to make an educated assumption here and declared a postal address, a country code, and an email address obligatory. In addition, we foresee the phone-number as an optional field. The name of the data protection officer, in turn, is obligatory only for Art. 30. %
It is therefore not an obligatory element for our language to be defined in subsequent steps. Regarding the formats, standardized representations exist and are therefore used for country codes, phone numbers, and email addresses. The remaining fields, in turn, are considered as simple strings.

\paragraph{Data Disclosed $\rightarrow$ Legal Basis:} The building block ``Data Disclosed'' integrates multiple transparency obligations related to particular categories of personal data collected and processed. For instance, Art. 13 and 14 both require to inform about ``the legal basis for the processing'' of personal data. This information has -- like those relating to other obligations in this building block -- to be provided in relation to specific categories of personal data. It must therefore be possible to repeat the whole ``Data Disclosed'' block multiple times, once for each category of personal data concerned.\footnote{See Data Disclosed $\rightarrow$ Data category.} The legitimating legal basis, in turn, can be a provision from the GDPR itself or from any other law or regulation such as a national disease control regulation, which, in some cases, may require additional remarks or explanations. We therefore deduce the need for a bipartite statement here, comprising an obligatory standardized reference to a legal provision and an optional free-text description.

In a similar vein, we analyzed all transparency obligations and specified required blocks for our transparency information language. A summary of the resulting blocks and data fields that need to be expressible is provided in table \ref{tab:building-blocks}. Notably, this also includes a separate building block with meta-information allowing to uniquely identify a particular record of transparency information, to specify validity dates, or to ensure its integrity. These allow for more advanced functionalities at later stages. 

\begin{savenotes}
\begin{table}[t!]
  \caption{Building blocks to be expressed within TILT}
  \label{tab:building-blocks}
\small 
\begin{multicols}{2}
\begin{tabular}[]{Hp{0.95\columnwidth}HH}
& \textbf{Meta}\\\midrule
	\multirow{9}{4cm}{Meta information for the identification and verification of the document} & 1. Identification Number\req\footnote{Follows the database-specific implementation; should offer as much entropy for globally unique identifiers.} & 1. f1424f86-ca0f-4f0c-9438-43cc00509931 & \multirow{2}{5cm}{1. The identification number is globally unique.} \\ 
	 & 2. Name\req & 2. GreenCompany &  \\ 
	 & 3. Creation date\req & 3. 2020-04-03T15:53:05.929588 &  \\ 
	 & 4. Modification date\req & 4. 2020-06-01T16:16:47.151300 &  \\ 
	 & 5. Version\req & 5. 2 &  \\ 
	 & 6. Language code\req\footnote{All language abbreviation codes follow the established ISO 639-1 standard as identifiers for names of languages.} & 6. de &  \\ 
	 & 7. Status\req & 7. active &  \\ 
	 & 8. URL\req & 8. https://green-bikes.de/privacy &  \\ 
	 & 9. Hash\req\footnote{The hash is based on one SHA256 calculation of the document content.} & 9. d732a793562a3e5dc57645a8e9\ldots &  \\ 
\\ & \textbf{Controller}\\\midrule
	\multirow{7}{4cm}{Responsible controller} & 1. Company name\req & 1. GreenCompany AG \\
	 & 2. Division\footnote{Differentiates internally of a company; relevant for large companies.} & 2. Product line e-mobility & \multirow{4}{5cm}{2. Serves to differentiate between different areas of a company; particularly relevant for large companies.} \\ 
	 & 3. Address\req & 3. Wolfsburger Ring 2, 38440 Berlin &  \\ 
	 & 4. Country code\req\footnote{All country codes follow the established ones ISO 3166 country abbreviation standard.} & 4. DE &  \\ 
	 & 5. Name (representative)\req & 5. Jane Super &  \\ 
	 & 6. Email (representative)\req & 6. contact@greencompany.de &  \\ 
	 & 7. Phone (representative) & 7. 0151 1234 5678 &  \\ 

\\ & \textbf{Data Protection Officer}\\\midrule
	\multirow{5}{4cm}{Data protection officer} & 1. Name & 1. Data protection officer of GreenCompany AG & \multirow{2}{5cm}{1. If possible, a natural person must be specified.} \\ 
	 & 2. Address\req & 2. Wolfsburger Ring 2, 38440 Berlin &  \\ 
	 & 3. Country code\req & 3. DE &  \\ 
	 & 4. Email\req & 4. privacy@greencompany.de &  \\ 
	 & 5. Phone & 5. 0171 1234 5678 &  \\

	 \\ & \textbf{Data Disclosed}, \textit{per item:}\\\midrule
	 \multirow{1}{4cm}{Data (categories)} & 1. Data category\req & 1. E-mail address
	 
	 & \multirow{10}{5cm}{1. The data category can be taken from a collection of standard vocabulary or, in special cases, chosen freely.\\3. The legal bases are based on the exact information from the respective law according to GDPR, BDSG etc. A legal basis is coded according to a format to be specified (see below).\\4. The legitimate interest only has to be stated if the processing is carried out in accordance with Art. 13 (1d).\\5. The recipient is represented in the same way as the controller if possible, otherwise only the category is given.} \\

	 \multirow{3}{4cm}{Purposes} & 2. [Purpose]\req & 2. [(\enquote{Marketing}, \enquote{Newsletter once a month})] &  \\ 
	 & \hspace*{0.5cm} a. Purpose\req &  &  \\ 
	 & \hspace*{0.5cm} b. Description\req & 3. [(GDPR-99-1-a, It will...), (BDSG-42-1)] &  \\ 
	 \multirow{3}{4cm}{Legal bases} & 3. [Legal basis]\req &  & \\ 
	 & \hspace*{0.5cm} a. Reference\req &  &  \\ 
	 & \hspace*{0.5cm} b. Description &  &  \\ 
	 \multirow{3}{4cm}{Legitimate interests} & 4. [Legitimate interest]\req\footnote{The legitimate interest only has to be stated if the processing is carried out in accordance with Art. 13 (1d).} & 4. [True, \enquote{There is an interest in\ldots}] &  \\ 
	 & \hspace*{0.5cm} a. Exists?\req &  &  \\ 
	 & \hspace*{0.5cm} b. Reasoning &  &  \\ 
	 \multirow{5}{4cm}{Recipients (categories)} & 5. [Recipient (category)]\req\footnote{Thus, any third country transfers are also defined.} & 5. [(GreenComp, Sales, 5th Avenue...), (...)] &  \\ 
	 & \hspace*{0.5cm} a.--d. \textit{(as controller)} &  &  \\ 
	 & \hspace*{0.5cm} e. Category\req &  &  \\ 

	 \multirow{11}{4cm}{Duration of storage or storage criteria} & 6. [Storage periods]\req & \multirow{3}{8cm}{6. [(2005-08-09T18:31:42P3Y6M4DT12H30M17S), (Until the end of the ordering process), (SGB-42), (min | max | sum | avg)]} & \multirow{5}{5cm}{6. The aggregation function describes the calculation basis when specifying several time intervals. For example, if there is storage for 2 weeks for technical reasons (e.g. backup), but there is a legally longer retention period, the maximum aggregation function (max) would be selected (standard case).} \\ 
	 & \hspace*{0.5cm} a. temporal\footnote{\enquote{A time period is represented in the format [start date]P[YY][MM][WW][TD] [T[hH][mM][s[.f]S]]. The \textit{P} indicates as a leading information letter that a period follows. Periods that contain a portion of the time are delimited by a T, as in the specification of the start time. It is therefore possible to distinguish between the months and minutes (M). The same rules apply for formatting the starting time as for normal dates.} (own translation) Cf. \url{https://de.wikipedia.org/wiki/ISO_8601\#Zeitspannen}. Note: In most cases the start date would probably be omitted or implicitly results from the time the personal data was collected.}: TTL\footnote{The TTL (\enquote{time to live}) specifies a specific \enquote{lifetime} in the form of a time span.} &  &  \\ 
	 & \hspace*{0.5cm} b. conditional: Purpose &  &  \\ 
	 & \hspace*{0.5cm} c. conditional: Legal basis &  &  \\ 
	 & \hspace*{0.5cm} d. Aggregation function\footnote{The aggregation function describes the calculation basis when specifying several time intervals. For example, if there is storage for 2 weeks for technical reasons (e.g. backup), but there is a legally longer retention period, the maximum aggregation function (max) would be selected (standard case).} &  &  \\

\\ & \textbf{Adequacy decisions}\\\midrule
	\multirow{16}{4cm}{Adequacy (third country)} & 1. Third country abbreviation & 1. Third country abbreviation per third country. & \multirow{3}{5cm}{The sample of the companies examined in most cases indicates the existence of a standard data protection clause.} \\ 
	 & 2. Decision by commission? & 2.--7. Binary decision (e.g. False) and a description as free text.  & \\ 
	 & 3. Appropriate guarantees?\footnote{According to Art. 45. The \enquote{suitable guarantees can be available without special authorization from a supervisory authority} (Art. 15) under certain conditions.} &  &  \\ 
	 & 4. Description of guarantees &  &  \\ 
	 & 5. Rights and effective remedies &  &  \\ 
	 & 6. Description thereof &  &  \\ 
	 & 7. Standard protection clause? &  &  \\ 
\end{tabular}

\begin{tabular}[]{Hp{0.95\columnwidth}HH}
	 & \textbf{Access/Data portability}\\\midrule
	\multirow{9}{4cm}{Access and data portability} & 1. Possibility available?\req & 1. True & \multirow{7}{5cm}{The information is subject to the requirements of Art. 20 (right to data portability). The possible administrative fee (for several copies) was not discernible in the cases examined so far.} \\ 
	 & 2. Description accessibility & 2. \enquote{In the event that the requirements of Art. 20 Para. 1 GDPR are met, you have the right to store your data in a structured, common ...} (cf. DHL; see below) &  \\ 
	 & 3. URL & 3. https://greencompany.de/access &  \\ 
	 & 4. E-mail & 4. access@greecompany.de &  \\ 
	 & 5. [Identification evidence] & 5. [ID card copy, email verification] &  \\ 
	 & 6. Administrative fee &  &  \\
	 & \hspace*{0.5cm} a. Amount  b. Currency & 0.00 &  \\ 
	 & 7. Data format(s) & 7. [JSON] & Commonly structured machine-readable formats are, for example, JSON, XML or YAML. \\

	 \\ & \textbf{Sources}\footnote{This duty to provide information is limited to the collection of personal data that does not take place from the data subject (Art. 14).}, \textit{per item:}\\\midrule

	 & 1. Data (category) & 1. Credit worthiness &  \\ 
	 & 2. [Sources] & 2. $\neg$ &  \\ 
	 & \hspace*{0.5cm} a. Description & \hspace*{0.5cm} a. Schafu &  \\ 
	 & \hspace*{0.5cm} b. URL &  \hspace*{0.5cm}b. https://schafu.de/ &  \\ 
	 & \hspace*{0.5cm} c. Publicly available? & \hspace*{0.5cm} c. False &  \\ 
	 
 \\ & \textbf{Rights to information,\newline rectification/deletion,\newline data portability,\newline withdraw consent}, \textit{for each}:\\\midrule
	\multirow{7}{4cm}{Right to information} & 1. Possibility available?\req & 1. True &  \\ 
	 & 2. Description & 2. For the right to information please use this contact form and ... & Standard texts could be specified.\\ 
	 & 3. URL & 3. https://greecompany.de/yourRights &  \\ 
	 & 4. E-mail & 4. - &  \\ 
	 & 5. [Identification evidence] & 5. [Censored photocopy of identity card, date of birth] &  Standard evidence could be provided \\

\\ & \textbf{Right to complain}\\\midrule
& 1.--5.: \textit{see above}\\
	 & 6. Supervisory authority\req & 6. Commissioner for Data Protection &  \\ 
& 7.--10.: \textit{contact details}

\\ & \textbf{Non-disclosure}\footnote{Refers to the necessity and consequences in the event of non-disclosure: According to Art. 13 (2e), this refers to the information whether the provision of the personal data is required by law or contract or is required for the conclusion of a contract, whether the data subject is obliged to provide the personal data and the possible consequences of not providing it.}, \textit{per item:}\\\midrule
	 & 1. Data (category)\req & 1. Contact details including email address &  \\ 
	 & 2. Legal requirement?\req & 2. False &  \\ 
	 & 3. Contractual regulation?\req & 3. True &  \\ 
	 & 4. Obligation to provide?\req & 4. True &  \\ 
	 & 5. Consequences if not provided\req & 5. The shipment cannot be delivered. &  \\ 

\\ & \textbf{Automated decision making}\\\midrule
	\multirow{5}{4cm}{Automated decision making and logic} & 1. In use?\req & 1. True & \multirow{3}{5cm}{The more detailed descriptions are given when automated decision making (including profiling) exists.} \\ 
	 & 2. Logic involved & 2. The personal data are processed as follows: ... &  \\ 
	 & 3. Scope and intended effects of processing for the data subject & 3. From processing follows ... &  \\ 

\\ & \textbf{Notification on change}\\\midrule
	\multirow{6}{4cm}{Notification of change of purpose} & 1. Affected categories & 1. A new data protection officer was hired ... & \multirow{4}{5cm}{The URL points to a new transparency policy implicitly creating a chain of transparency information documents.} \\ 
	 & 2. Date time of change & 2. 2020-08-2020-06-04T15:04:13+0000 &  \\ 
	 & 3. URL of last document & 3. https://greencomp.de/privacypolicy/2 &  \\
\end{tabular}
\newline
\newline
\newline\flushleft{\textbf{Legend:}\\ \textit{Item}\req: Mandatory information\newline [\textit{Item}]: List (possibly empty)}
\newline
\end{multicols}

\end{table}

\end{savenotes} %

\section{Language Design and Implementation}
\label{sec:language design and implementation}

Based on the above analysis of the required expressiveness and the identification of data formats to be used, we designed our transparency information language (TIL), following well-established steps from language design \cite{whenAndHow2005} also employed for other privacy-related languages \cite[cf. ][]{UlbrichtYaPPLLightweightPrivacy2018}. In particular, this includes a complete formal grammar in Extended Backus-Naur form (EBNF) \cite{standard1996ebnf} and a complete JSON\footnote{For the choice of JSON over other formats, see below.} Schema \textit{v7} specification \cite{jsonschema2016}. Both are, again due to space constraints, available online.\footnote{\url{https://github.com/Transparency-Information-Language/schema}} Some core concepts shall, however, be briefly elucidated below.

\subsection{Formal definition}
Serving as a reference for different implementations, 
the formal definition of our language translates each of the building blocks from table~\ref{tab:building-blocks} into a finite set of production rules. In anticipation of the JSON Schema implementation, some extra fields are necessary to that end. The complete set of productions creates a context-free grammar. 

First (cf. listing~\ref{lst:grammar1}), the root element \lstinline[basicstyle=\ttfamily\color{black}]|tilt| consists of three associated non-terminal symbols which are \lstinline[basicstyle=\ttfamily\color{black}]|header| and \lstinline[basicstyle=\ttfamily\color{black}]|properties|, and an optional \lstinline[basicstyle=\ttfamily\color{black}]|[addProp]|. The latter allows any customization and extension of the language for their users if needed (see \textit{Req. 4}). After having defined the root element and associated non-terminals, all \lstinline[basicstyle=\ttfamily\color{black}]|property| non-terminals are resolved to either their relative non-terminal symbols or directly to basic types, which are described at the end of the specification. The \lstinline[basicstyle=\ttfamily\color{black}]|meta| non-terminal field \lstinline[basicstyle=\ttfamily\color{black}]|id|, for instance, directly resolves to a \lstinline[basicstyle=\ttfamily\color{black}]|string| value which is a basic type. All other non-terminals describe the individual fields of the building blocks listed in table~\ref{tab:building-blocks}.

\vspace{\baselineskip}
\begin{lstlisting}[style=ebnf, label=lst:grammar1, firstnumber=16, basicstyle=\ttfamily\footnotesize, caption={Grammar for root element, JSON Schema meta fields, and main properties}]
tilt = header, properties, [addProp];

header = schema, id, title, description, examples, [addProp];
schema = 'http://json-schema.org/draft-07/schema';
id = 'https://github.com/<anonymized>';
title = 'Transparency Information Language';
description = string;
examples = { properties };

properties = meta,
  controller, dataProtectionOfficer, dataDisclosed,
  thirdCountryTransfers, accessAndDataPortability, sources,
  rightToInformation, rightToRectificationOrDeletion,
  rightToDataPortability, rightToWithdrawConsent, rightToComplain,
  automatedDecisionMaking, changesOfPurpose, [addProp];
\end{lstlisting}

\begin{figure}[!htb]
\noindent\begin{minipage}{.4\columnwidth}
	\begin{lstlisting}[style=ebnf, label=lst:grammar2, firstnumber=64, basicstyle=\ttfamily\footnotesize, caption={Grammar}]{}
dataDisclosed = {(
	id,
	category,
	purposes,
	legalBases,
	legitimateInterests,
	(recipients | category),
	storage,
	nonDisclosure,
	[addProp]
)};

category = string;
purposes =
	 (purpose, description)
	 | purposes;
purpose = string;
...
	\end{lstlisting}
\end{minipage}\hfill
\begin{minipage}{.6\columnwidth}
	\centering
		\vspace*{0.4cm}
	\includegraphics[width=1.0\linewidth]{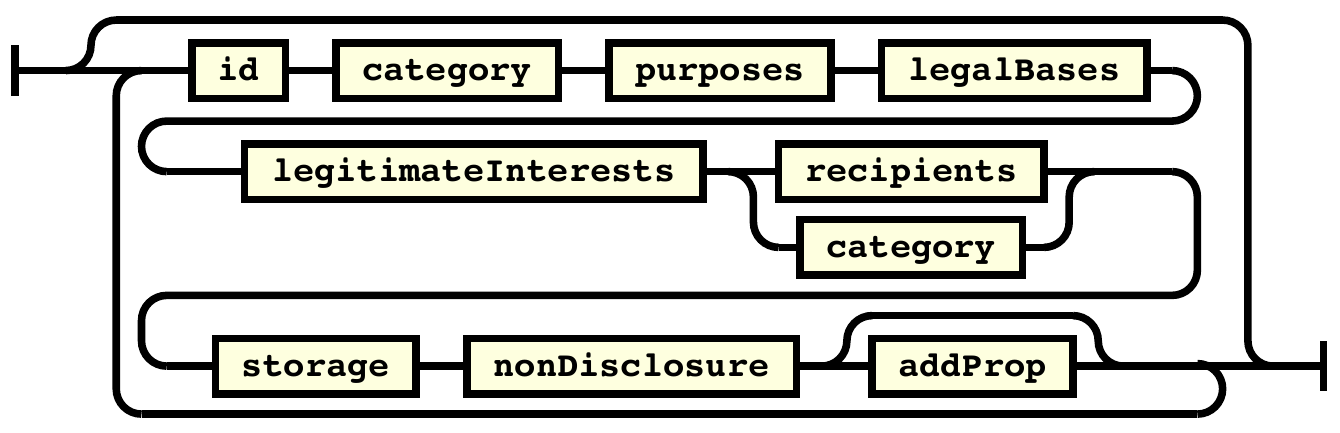}\\
		\vspace*{2.30cm}
	\includegraphics[width=.67\linewidth]{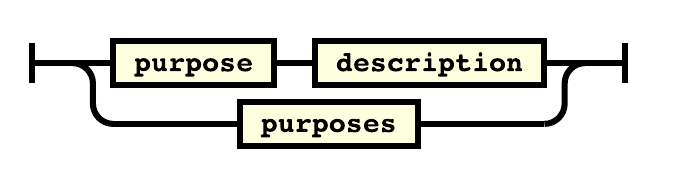}\\
		\vspace*{0.4cm}
	\caption{Syntax diagrams}\label{fig:syntax}
\end{minipage}
\end{figure}

An exemplary building block is formally defined in listing~\ref{lst:grammar2} and visualized in figure~\ref{fig:syntax}: The \lstinline[basicstyle=\ttfamily\color{black}]|dataDisclosed| property definition reflects the cases according to Art. 13(1f) GDPR in which the recipients or categories thereof need to be specified. In many cases, the non-terminal symbols do not directly resolve to terminals as in the case of \lstinline[basicstyle=\ttfamily\color{black}]|purposes|: The additional production rules allow the recursive definition of a set of \lstinline[basicstyle=\ttfamily\color{black}]|(purpose, description)| pairs.

All fields are syntactically defined, not necessarily semantically. However, many fields carry implicit semantics in their syntax: The \lstinline[basicstyle=\ttfamily\color{black}]|stringDate| or \lstinline[basicstyle=\ttfamily\color{black}]|stringEmail| (not shown) fields are such examples for which the exact format does not allow semantically different inputs. These specifications are of particular relevance when technically implementing the language because the meta language at least needs to support data validation via regular expressions.

Finally, the specification comprises all basic types which resolve to terminal symbols. These include \lstinline[basicstyle=\ttfamily\color{black}]|string|, \lstinline[basicstyle=\ttfamily\color{black}]|boolean|, and \lstinline[basicstyle=\ttfamily\color{black}]|number| values. For many of them regular expressions (or type descriptions referring to regular expressions stated in there) are given. Since JSON is used as the meta language of expression, its specification is directly taken and adapted from the official grammar.\footnote{\url{https://www.json.org/json-en.html}}
We provide a visualized version of the grammar including all syntax diagrams online.

\subsection{Technical implementation}
One of the main contributions of this paper is the full language implementation in the form of a JSON Schema. Alternative technical implementations could also be developed with XML/XSD or RDF/RDFS. For reasons similar to the ones listed in \cite{UlbrichtYaPPLLightweightPrivacy2018} -- human readability, robustness, broad programming language and tooling support, and resource efficiency -- we opted for JSON/JSON Schema as the default representation format.\footnote{Other representations, particularly in XML, might be added later with reasonable effort. First steps in that direction are experimentally documented: \url{https://github.com/Transparency-Information-Language/tilt-experimental/tree/master/xsd}} In the following, we first present the core language implementation. In a second step, we then demonstrate how this implementation can be extended.

\subsubsection{Core language}
Regarding the core language model, in 2692 lines of code we represent every single field of all introduced building blocks. Each one has its own type definition, including a comprehensible description, at least one example value and in most cases further type specifications through regular expressions or other means in JSON schema. The schema also contains a complete document covering most of the possible paths of the syntax tree. Some properties, such as \texttt{\$schema}, \texttt{\$id}, \texttt{title} etc., are (de-facto\footnote{The minimal valid JSON Schema document is an empty object type (trivial).}) mandatory JSON Schema meta properties which are used to describe the document for validators. Afterwards, the language definition according to the formal definition is carried out. Moreover, all building blocks are marked as required since all of these properties have to be expressed explicitly (addressing \textit{Req. 1} and \textit{Req. 2}). The language allows additional properties as pointed out in the formal definition above, which is relevant for controllers that want to extend the schema by their needs apart from GDPR transparency information (\textit{Req. 4}). 

An example for custom type validation is the reference to a legal basis for data processing. Several conventions were introduced (see table~\ref{tab:building-blocks}) that need to be enforced within the technical implementation. In this example, the convention is materialized using a regular expression and its description as depicted in listing~\ref{lst:s5}. Likewise, valid and invalid examples are given in listings~\ref{lst:s6} and~\ref{lst:s7}. The first example successfully validates against the regular expression, whereas the second does not. JSON Schema validators\footnote{e.g. \url{https://github.com/ssilverman/snowy-json}} %
quickly reveal erroneous fields to be corrected.

\vspace{\baselineskip}
\begin{lstlisting}[language=json, label=lst:s5, firstnumber=846, basicstyle=\ttfamily\footnotesize, caption=dataDisclosed/legalBases/reference schema]
"properties": {
	"reference": {
		"$id": "#/properties/dataDisclosed/items/anyOf/0/properties/legalBases/items/anyOf/0/properties/reference",
		"type": "string",
		"title": "Reference",
		"description": "This field refers to the reference in legal regulations (laws, orders, declaration etc.). The format is set to uppercase letters for the legal text followed by hyphened numbers and lowercase letters for the exact location.",
		"pattern": "^[A-Z]*([-]?[0-9]*|[a-z]*)*$",
		"examples": [ "GDPR-99-1-a" ]
	},
	...
\end{lstlisting}

\begin{minipage}{.41\columnwidth}
\begin{lstlisting}[language=json, label=lst:s6, firstnumber=41, basicstyle=\ttfamily\footnotesize, caption=Valid legal basis]
"legalBases": [
 {
  "reference": "GDPR-6-1-a",
   "description": "The data subject has given..."
 },
  ...
\end{lstlisting}
\end{minipage}\hfill
\begin{minipage}{.41\columnwidth}
\begin{lstlisting}[language=json, label=lst:s7, firstnumber=41, basicstyle=\ttfamily\footnotesize, caption=Invalid legal basis]
"legalBases": [
 {
  "reference": "CCPA 1798",
  "description": "A consumer shall..."
 },
  ...
\end{lstlisting}
\end{minipage}
\vspace{\baselineskip}

Similar to the above-mentioned mechanisms, other fields are also validated. For instance, timestamps, language and country codes or international telephone numbers are validated using custom regular expressions according to the conventions stated above. JSON Schema also has built-in types, which are used through the \texttt{format} keyword. For instance, the \texttt{uri-reference} format is used to validate HTTP links to the controller's web page or the \texttt{accessAndDataPortability} measures.

Besides the validation of singular fields within (sub)schemas of properties, overarching interrelationships are validated by our language implementation as well. As laid out in Art. 22 GDPR %
and the Recitals 71, 72 and 91, data subjects needs to be informed about the logic involved in case automated decision making is in use. Consequently, there is an interdependency between existing fields within the language schema. Such requirements are solved using logical operators within JSON Schema. An \texttt{if-then} construct enforces the appearance of \texttt{logicInvolved} and \texttt{scopeAndIntendedEffects}, whenever the (automatedDecisionMaking) \texttt{inUse} property is set to \texttt{true}.

Another example for higher-order capabilities of our language is the GDPR requirement to state the \enquote{the period for which the personal data will be stored, or [\ldots] the criteria used to determine that period} according to Art. 13 (2a). Taking into account the requirements from real-world scenarios, different modalities to fulfill this requirement are apparent. The formal definition sets up three possibilities to define the time period or respective criteria. In the technical implementation the choice is constructed using an \texttt{anyOf} environment. The logic is implemented as part of the \texttt{storage} item. Consequently, controllers can express \texttt{temporal} as well as \texttt{purposeConditional} and \texttt{legalBasisConditional} options. 

Building domain-specific languages is an \enquote{incremental, modular, and extensible way from parameterized building blocks} as \cite{whenAndHow2005} state. The building blocks as worked out in table~\ref{tab:building-blocks} might be extended or changed over time, since a new iteration of GDPR is about to come. Therefore, the introduced language model has to be seen as a \enquote{living} specification that will have to adapt to new legislation over time.

In order to motivate the usage of our language in practical privacy engineering, we demonstrate several components of a toolkit that has been specifically designed around the technical implementation at hand in section \ref{sec:toolkit}. Before that, we illustrate how the language capabilities can be customized for a specific application domain in the following section. 

\subsubsection{Extension through vocabularies}

In many plausible cases, it should be possible to extend the expressiveness of the language. Reasons for this can be controller-specific requirements or the binding to other language definitions. Therefore we show how to specify single fields by using the example of purpose definitions. In \cite{pbac2020} the authors require purpose definitions to be hierarchical and based on flexible vocabularies with allowed and prohibited values. %
In the context of personal data sharing, different purposes may define \enquote{broad} or \enquote{narrow} consent respectively for which transparency should be guaranteed. Examples for such hierarchies are the following:

\begin{itemize}
	\item Research / Marketing / ... \textit{(broad)}
	\item Clinical research / Advertising / ... \textit{(rather broad)}
	\item COVID19 research / Targeted advertising / ... \textit{(rather narrow)}
	\item Polymerase chain reaction testing / Tracking technologies including mouse movements / ... \textit{(narrow)}
\end{itemize}

Our transparency information language can be extended at attribute level. Given the \texttt{purposeConditional} field, we can deviate from the simple \texttt{string array} and formulate a custom vocabulary in three steps\footnote{For a detailed description, refer to \url{http://github.com/Transparency-Information-Language/vocabularies}}:

First, we create a JSON schema for the new purpose field specifying all values and necessary rules as a stand-alone document. Using the \texttt{allOf}, \texttt{anyOf}, and \texttt{not} keywords every combination of purposes can be formulated. There is full flexibility on the purpose vocabulary. An accepted authority could publish a full definition of purposes under an open content license. Then, industry can adopt the vocabulary to incorporate new transparency standards. Adding the \texttt{enum}, \texttt{const} and \texttt{not} meta language elements, there can even be made a distinction between allowed and prohibited purposes. Secondly, we change the items specification within the original schema to the uri-reference where our new purpose vocabulary file can be found (online). Finally, we can validate the extended schema using existing schema validators.

As a consequence, this extension shows how the language definition can become more expressive using externally defined schemata. Therefore even full complementary privacy preference languages can be integrated. If another language is also implemented in JSON schema, data controllers can profit from the capabilities of both languages at the same time (\textit{Reqs. 4, 5, 6}). %

\section{Toolkit}
\label{sec:toolkit}

To meet the requirements for re-usability (\textit{Req. 5}) and developer-friendliness (\textit{Req. 6}), we also provide %
companion libraries for two programming languages widely used in the domain of (web) information systems: Java and Python. These libraries allow to programmatically create, manage, and validate transparency information objects. In addition, we crafted a fully implemented document storage with an easily usable remote API, providing inter-system communication and versioned persistence (\textit{Req. 3}) and heightening re-usability (\textit{Req. 5}) and developer-friendliness (\textit{Req. 6}) even further.

\subsection{General Purpose Programming Language Bindings}
With the above language alone, transparency information now can be expressed in a standardized manner using the JSON Schema implementation. The manual creation of these JSON documents can be supported by (visual) editors.\footnote{\url{https://jsoneditoronline.org/}} However, a manual workflow is error-prone and difficult to distribute among several people or departments in a company. Therefore, we provide two general purpose programming language bindings that allow to express the transparency information building blocks as native language elements.\footnote{\url{https://github.com/Transparency-Information-Language/java-tilt}}\textsuperscript{,}\footnote{\url{https://github.com/Transparency-Information-Language/python-tilt}} After the creation of such elements, these can be serialized as JSON documents according to the schema. Moreover, already existing JSON documents that successfully validate against the schema can be imported into the programming language at hand. Here, we show two implementations in Java and Python, demonstrating that and how transparency information can also be generated in a developer-centric manner in program code. This approach allows the involvement of the developers who ultimately implement the actual data processing and who know best which data is processed where and for which purpose. Other language bindings are planned to complement the toolkit as future work (candidate languages are JavaScript, Kotlin, Swift etc.) for covering a more diverse field of application development.\footnote{The \url{quicktype.io} project helped us to create the prototypes.}

The Java implementation of our language allows the dynamic creation of JSON documents according to the language schema. The core consists of 34 classes. For each class, the fields are only accessible through \textit{Getter} and \textit{Setter} methods. Each field is annotated using the \texttt{jackson}\footnote{\url{https://github.com/FasterXML/jackson}} library for efficient (de)serialization. Additional \lstinline[language=java]|toString()| methods help for testing and logging purposes. %
Listing~\ref{lst:java} represents the low-effort use of the implementation (\textit{Req. 6}) in a shortened form: \textit{(1)} The central \texttt{Tilt} object type encapsulates all pieces of transparency information which are empty by default. \textit{(2)} For the validation of external documents only the URL has to be provided. \textit{(3)} Existing documents can be de-serialized and manipulated using the \texttt{Converter} class.
For seamless distribution, the software project is automatically packaged using the Github automation workflow. As a result, the package is available through popular dependency repositories by adding the reference to our implementation to the traditional \texttt{pom.xml} file (\textit{Req. 5}).
We emphasize that by providing the package, it is now possible to integrate the language into the context of large distributed systems. Java is particularly popular in the context of service-oriented architectures and data-intensive web services \cite{Wetherbee2018}. Modern microservice architectures build upon loosely coupled message oriented communication models (\textit{Req. 3}). Through the included JSON adapters, the integration should be realizable with little development efforts (\textit{Req. 6}).\\

\vspace{\baselineskip}
\begin{lstlisting}[language=java, label=lst:java, firstnumber=1, basicstyle=\ttfamily\footnotesize, caption={Java language binding example.}]
// 1 - Generate new Tilt instance from scratch
Tilt tilt = new Tilt();
Controller controller = new Controller();
controller.setName("Example Company SE");
tilt.setController(controller);
// 2 - Validate existing documents
TiltValidator.validateDocumentFromUrl(DOCUMENT_URL);
// 3 - De-serialize and manipulate existing documents
Tilt t = Converter.fromJsonString(instance);
t.getMeta().setHash("42");
\end{lstlisting}

The Python library functions very similar to the previously described implementation. It consists of class representations and several \texttt{$\_\_$init$\_\_$(args)} and helper functions for type checking and conversion. With the exception of the \texttt{json} library, there are no other dependencies, which makes the full language binding uncomplicated and easy to integrate into existing software projects (\textit{Req. 5}). Along with the core module comes the documentation of the open source project and several example programs that are linked to an interactive online playground. The library, documentation, etc. are provided as a PyPI package.\footnote{\url{https://pypi.org/project/tilt/}} %
Therefore, the tool can be installed by every developer using the standard package sources.\footnote{\texttt{pip install \textanon{tilt}{<anonymized>}}} For the validation tasks, we have successfully tested and therefore recommend the \texttt{fastjsonschema}\footnote{\url{https://pypi.org/project/fastjsonschema/}} library.

Both language bindings have in common that they facilitate the adaptive creation of transparency information documents. Having the tool at hand allows developers to manifest the transparency notice in source code on a par with the relevant data processing logic (\textit{Req. 6}).

\subsection{Document storage with Application Programming Interfaces}\label{sec:tilt-hub}

In order to complement the toolkit and for exploring possible storage and interaction models with regards to machine-readable transparency information, in this section the component \textit{tilt-hub}, an extensible document storage, is introduced.\footnote{\url{https://github.com/Transparency-Information-Language/tilt-hub}} Within \textit{tilt-hub}, documents expressed in our language can be stored and versioned.\footnote{For this purpose, we introduced the \textit{meta} building block containing modification date, status, hash, and identification number. These fields can express the timeliness, sequence, integrity and status of documents.} Furthermore, two APIs enable 1) queries to retrieve the exact transparency information that is needed for a specific use case and 2) queries to reveal data sharing networks if the documents are describing several data processing entities.%

\begin{figure}[!htb]
	\centering
	\includegraphics[width=0.7\linewidth]{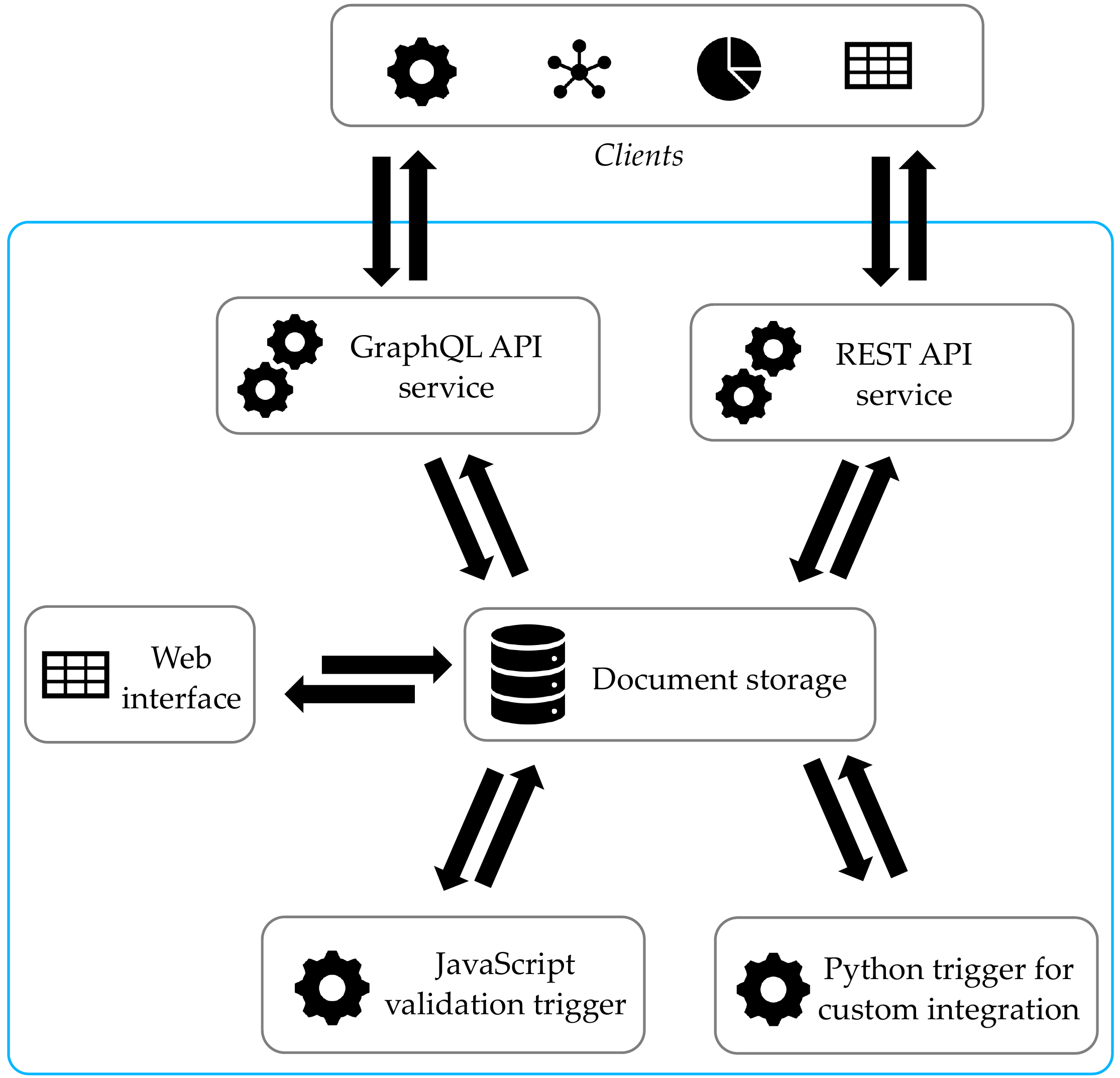}
	\caption{System architecture of the document storage.}
	\label{fig:tilt-hub-architecture}
\end{figure}

Figure \ref{fig:tilt-hub-architecture} shows the system architecture which relies on the concept of microservices. Each implemented service can be scaled up and down individually using docker-compose.\footnote{\url{https://docs.docker.com/compose/}} The central component is an instance of MongoDB\footnote{\url{https://www.mongodb.com}}, a JSON based document database, per default running as a single node cluster. It could be configured to run on horizontally scaled infrastructure to be able to store millions of transparency-related documents. Since MongoDB is published under an open source license, a big ecosystem of various toolkits, APIs and GUIs already has evolved. %
In particular, this includes the Mongo Express\footnote{\url{https://github.com/mongo-express/mongo-express}} web-based user interface %
and driver support for almost all popular programming languages.\footnote{\url{https://docs.mongodb.com/drivers/}} %
In addition to the aforementioned native drivers for MongoDB, two language-agnostic application programming interfaces were developed for the document storage, namely a GraphQL API service and a REST API service. Both are realized as independently functioning microservices. By the help of these services, clients can for instance perform search queries, integrate the transparency information language into their continuous integration or deployment pipelines, or connect other privacy-related software components such as personal information management systems. When building applications, these techniques and the developed language bindings play hand in hand to lower implementation costs for the efficient storage and queries of transparency documents (\textit{Reqs. 3 and 6}). To the best of our knowledge, these are distinguishing features, not entirely present in related work.

Furthermore, the infrastructure comprises two additional \textit{trigger}ing microservices. Both of them show the database interaction within the internal \textit{tilt-hub} network. The Python trigger serves as a starting point for developers who want to integrate the document storage into their CI/CD pipelines. An exemplary use case might be the notification in case of changes: A developer could have implemented a new function in one of the data controller's services that introduces a new third country transfer. Consequently, the developer pushes the change to \textit{tilt-hub} e.g. using the REST API. This upsert would trigger the Python microservice to send an email or short-message notification to the law department informing about the successful update. The Python microservice uses the native language binding described above and connects to the database using pymongo.\footnote{\url{https://pypi.org/project/pymongo/}} Additionally, a JavaScript microservice using node.js\footnote{\url{https://nodejs.org/en/}} was implemented for the continuous validation of all changes to the database with respect to our JSON schema. This microservice had to be developed since the integrated schema validation capabilities of MongoDB are limited to version 4 of the JSON schema reference implementation\footnote{\url{https://docs.mongodb.com/manual/core/schema-validation/}} while our language relies on the more recent version 7. For the schema validation we employ the node.js library Ajv that has been benchmarked as the fastest validator available for JavaScript.\footnote{\url{https://github.com/ajv-validator/ajv}}
We propose the usage of \textit{tilt-hub} for \textit{(i)} internal deployments only within the control sphere of a data controller for better internal document management, or \textit{(ii)} also with external access for interested data subjects and third parties. Moreover, \textit{(iii)} data protection officers or supervisory authorities could use the interfaces to interact with the transparency document storage for their tasks. Consequently, different deployment models offer a multitude of options for business process optimizations because they replace manual information management and analysis tasks (also addressing \textit{Req. 2}).\footnote{\url{https://github.com/Transparency-Information-Language/tilt-sample-analysis}}

\section{Exemplary Applications}
\label{sec:exemplary}

To demonstrate the opportunities for data subjects' informedness and data sovereignty unlocked by the existence of a structured, machine-readable language and respective tools as introduced above, we implemented two exemplary applications utilizing different capabilities and characteristics of these artifacts: A transparency analysis platform allowing to analyze stated data usage and sharing practices across multiple data processing parties and a simple browser extension illustrating the possibilities for novel user-facing representations of transparency information (cf. \textit{Req. 2}).

\subsection{Transparency Analysis Platform}\label{sec:tap}

We built a transparency analysis platform that serves as a tool for inspection and analysis tasks related to the presented transparency information. With this tool, it is possible to map respective information of a multitude of data controllers onto a graph structure. This allows to reveal potential data sharing networks among multiple data controllers that are hard or impossible to grasp by individual data subjects today as doing so would require scanning thousands of traditional privacy policies and putting them into context. In order to make use of publicly available transparency information from existing transparency and consent frameworks such as IAB Europe, we provide this ready-to-use analysis tool.

An already established software stack for graph databases serves as technical basis.\footnote{\url{https://grandstack.io/}} The prototypical implementation can be subdivided into data extractors and data processing components (Neo4j graph database, GraphQL API, and accompanying libraries). As an input, our platform accepts documents in the two different versions of the IAB Europe framework as well as documents expressed in our proposed language. The output is a visual graph structure or textual query results depending on the interfaces that are used.

With our tool, we were able to combine transparency information documents from different sources and explore complex controller-processor-purpose interrelations. These e.g. include \enquote{isolated}, \enquote{networked} or \enquote{linked} controller relationships which, for instance, may point at unforeseen knowledge concentrations or risks of interlinking data from different sources. %
Future work on the analysis platform includes pre-defined queries that answer specific questions well-known from social network or graph analysis studies \cite{knoke2019social, warchal2012using, rossi}. For now, we prove the applicability and interoperability of our language model within a practical privacy engineering software artifact.\footnote{\url{https://github.com/Transparency-Information-Language/transparency-analysis-platform}}

\subsection{Internet Browser Extension}
At European level, the Art. 29 WP \cite{wp29} recommended \enquote{that layered privacy statements/notices should be used %
[\ldots] rather than displaying all such information in a single notice on the screen} (cf. \textit{Req. 1}). Hence our toolkit and applicability examples are augmented by the development of an internet browser extension that is able to summarize key aspects of the transparency information expressed in our language. 

We target the Google Chrome browser platform to reach the large majority of internet users.\footnote{https://gs.statcounter.com/browser-market-share} Therefore, the architecture of the browser extension follows the developer guidelines provided by Google and is built upon web technologies including HTML, CSS, and JavaScript. The core functionality of the browser extension lies in fetching the language schema and the document expressed in our transparency language from publicly available sources (i.e. Github or the data controller's web server). Then, these contents are parsed and transformed into a summarizing textual and visual representation. For example, the stated data transfers are counted, the indication regarding automatic decision making is color-coded or the data protection officer is named. Third country transfers are represented using the official flags according to the country codes. The information is updated at each click on the extension's icon to prevent unnecessary HTTP requests. 

For future work, the browser extension should provide users with more diverse and dynamic representations according to their individual preferences and competencies. The inclusion of privacy icons, other visual-based representations or helping comments hold a lot of promise for adaptive representations \cite{holtz, Holtz2011}. Moreover, there is a direct link to the \textit{tilt-hub}, introduced in section~\ref{sec:tilt-hub}. As the different deployment models of \textit{tilt-hub} allow to provide transparency information independently from every single data controller and potentially even from a public repository, the browser extension could even display a summary of the transparency notice without the original publication of transparency information by the vendor under study. 

\section{Future work}
\label{sec:discussion}

The scientific discourse on privacy languages has been ongoing for several years already. Our language has a clear and in-depth focus on current transparency requirements from the GDPR. However, the language design and implementation process is meant to be adjusted to the technical and legal circumstances of tomorrow as well. Consequently, we see our language model as \enquote{living} specification that can and shall be extended by experts from various disciplines. With respect to traditional language evaluation criteria, we assume broad conformance \cite{paige2000principles}. However, to support the application in a diverse field of domains and for different target groups (from controller to data subject), we consider at least the following concepts worth investigating in future work. First, the automated extraction of transparency information from natural language or even source code (annotations) seems promising. In addition, there is a large body of literature on explainable AI which clearly relates to the information obligations with regards to automatic decision making \cite{ai}. Moreover, the integration and interplay with other privacy languages, e.g. in the context of legally sufficient consent \cite{UlbrichtYaPPLLightweightPrivacy2018}, needs to be examined. Later on, advanced concepts from the GDPR (e.g. joint controllership) or recent judicial decisions (e.g. relating EU--US data transfers) need to be taken into consideration. Finally, we also mentioned overlaps with accountability-related requirements for keeping records of processing activities from Art. 30 of the GDPR. These should also be investigated further in the future.

\section{Conclusion}
\label{sec:conclusion}

The current practice of providing privacy-related transparency information is dysfunctional and does not serve its originally intended goals anymore. To pave the way for novel, technically mediated approaches to transparency, we herein presented a structured, machine-readable transparency information language, accompanied by a surrounding toolkit that eases practical adoption. The language specification is based on an extensive analysis of transparency requirements from the GDPR and thus offers the expressiveness required to unfold actual legal relevance. The toolkit, in turn, particularly comprises two libraries for widely used programming languages as well as an easily instantiatable storage and management backend that significantly reduce practical implementation efforts for employing our language. Together, language and toolkit thus provide viable technical means for representing transparency information in a structured, machine-readable form as well as for managing and processing them automatically in real-world (web) information systems with reasonable implementation effort. %

Altogether, our language and toolkit are powerful building blocks allowing for the development of a broad variety of novel technical means for materializing the privacy principle of transparency in a way that better aligns with the original goals of respective regulations than current, legalese policy documents do. The transparency analysis platform and the browser extension presented herein vividly demonstrate these capabilities and we envision a broad range of further possible applications and tools to become possible on top of an existing machine-readable representation of transparency information.

Finally, by providing our transparency information language and toolkit, we also change the given preconditions for applying regulatory provisions: Art. 25 of the GDPR obligates the implementation of technical measures for \emph{all} privacy principles as soon as they are available as part of the current state of the art and can be implemented with reasonable effort. With the contributions presented herein, both factors come closer to the point where technical transparency mechanisms become obligatory.%

Insofar, our approach addresses the current structural weaknesses of transparency by taking an interdisciplinary point of view between law and technology and materializing the concepts into practical systems engineering. In so doing, we hope to re-amplify transparency that \enquote{might lead consumers to behave differently} and, thus, to contribute to an upcoming \enquote{renaissance} in privacy law \cite{kaminski2020recent}.

\begin{acks}
    The work behind this paper was partially conducted within the project DaSKITA, supported under grant no. 28V2307A19 by funds of the Federal Ministry of Justice and Consumer Protection (BMJV) based on a decision of the Parliament of the Federal Republic of Germany via the Federal Office for Agriculture and Food (BLE) under the innovation support programme.
\end{acks}

\bibliographystyle{ACM-Reference-Format}
\bibliography{bibliography} 

%%% -*-BibTeX-*-
%%% Do NOT edit. File created by BibTeX with style
%%% ACM-Reference-Format-Journals [18-Jan-2012].

\begin{thebibliography}{42}

%%% ====================================================================
%%% NOTE TO THE USER: you can override these defaults by providing
%%% customized versions of any of these macros before the \bibliography
%%% command.  Each of them MUST provide its own final punctuation,
%%% except for \shownote{}, \showDOI{}, and \showURL{}.  The latter two
%%% do not use final punctuation, in order to avoid confusing it with
%%% the Web address.
%%%
%%% To suppress output of a particular field, define its macro to expand
%%% to an empty string, or better, \unskip, like this:
%%%
%%% \newcommand{\showDOI}[1]{\unskip}   % LaTeX syntax
%%%
%%% \def \showDOI #1{\unskip}           % plain TeX syntax
%%%
%%% ====================================================================

\ifx \showCODEN    \undefined \def \showCODEN     #1{\unskip}     \fi
\ifx \showDOI      \undefined \def \showDOI       #1{#1}\fi
\ifx \showISBNx    \undefined \def \showISBNx     #1{\unskip}     \fi
\ifx \showISBNxiii \undefined \def \showISBNxiii  #1{\unskip}     \fi
\ifx \showISSN     \undefined \def \showISSN      #1{\unskip}     \fi
\ifx \showLCCN     \undefined \def \showLCCN      #1{\unskip}     \fi
\ifx \shownote     \undefined \def \shownote      #1{#1}          \fi
\ifx \showarticletitle \undefined \def \showarticletitle #1{#1}   \fi
\ifx \showURL      \undefined \def \showURL       {\relax}        \fi
% The following commands are used for tagged output and should be
% invisible to TeX
\providecommand\bibfield[2]{#2}
\providecommand\bibinfo[2]{#2}
\providecommand\natexlab[1]{#1}
\providecommand\showeprint[2][]{arXiv:#2}

\bibitem[\protect\citeauthoryear{Angulo, Fischer-H\"{u}bner, Pulls, and
  W\"{a}stlund}{Angulo et~al\mbox{.}}{2015}]%
        {angulo2015}
\bibfield{author}{\bibinfo{person}{Julio Angulo}, \bibinfo{person}{Simone
  Fischer-H\"{u}bner}, \bibinfo{person}{Tobias Pulls}, {and}
  \bibinfo{person}{Erik W\"{a}stlund}.} \bibinfo{year}{2015}\natexlab{}.
\newblock \showarticletitle{Usable Transparency with the Data Track: A Tool for
  Visualizing Data Disclosures}. In \bibinfo{booktitle}{\emph{Proceedings of
  the 33rd Annual ACM Conference Extended Abstracts on Human Factors in
  Computing Systems}}. \bibinfo{publisher}{ACM}, \bibinfo{address}{New York,
  NY, USA}, \bibinfo{pages}{1803–1808}.
\newblock
\showISBNx{9781450331463}
\urldef\tempurl%
\url{https://doi.org/10.1145/2702613.2732701}
\showDOI{\tempurl}


\bibitem[\protect\citeauthoryear{{Article 29 Data Protection Working
  Party}}{{Article 29 Data Protection Working Party}}{2017}]%
        {wp29}
\bibfield{author}{\bibinfo{person}{{Article 29 Data Protection Working
  Party}}.} \bibinfo{year}{2017}\natexlab{}.
\newblock \bibinfo{booktitle}{\emph{Guidelines on Transparency under Regulation
  2016/679}}.
\newblock \bibinfo{type}{{T}echnical {R}eport}. \bibinfo{institution}{Directive
  95/46/EC of the European Parliament}.
\newblock


\bibitem[\protect\citeauthoryear{Bhatt, Xiang, Sharma, Weller, Taly, Jia,
  Ghosh, Puri, Moura, and Eckersley}{Bhatt et~al\mbox{.}}{2020}]%
        {ai}
\bibfield{author}{\bibinfo{person}{Umang Bhatt}, \bibinfo{person}{Alice Xiang},
  \bibinfo{person}{Shubham Sharma}, \bibinfo{person}{Adrian Weller},
  \bibinfo{person}{Ankur Taly}, \bibinfo{person}{Yunhan Jia},
  \bibinfo{person}{Joydeep Ghosh}, \bibinfo{person}{Ruchir Puri},
  \bibinfo{person}{Jos\'{e} M.~F. Moura}, {and} \bibinfo{person}{Peter
  Eckersley}.} \bibinfo{year}{2020}\natexlab{}.
\newblock \showarticletitle{Explainable Machine Learning in Deployment}. In
  \bibinfo{booktitle}{\emph{Proceedings of the 2020 Conference on Fairness,
  Accountability, and Transparency}} \emph{(\bibinfo{series}{FAT* '20})}.
  \bibinfo{publisher}{Association for Computing Machinery},
  \bibinfo{address}{New York, NY, USA}, \bibinfo{pages}{648–657}.
\newblock
\showISBNx{9781450369367}
\urldef\tempurl%
\url{https://doi.org/10.1145/3351095.3375624}
\showDOI{\tempurl}


\bibitem[\protect\citeauthoryear{Bier, K{\"{u}}hne, and Beyerer}{Bier
  et~al\mbox{.}}{2016}]%
        {bier2016}
\bibfield{author}{\bibinfo{person}{Christoph Bier}, \bibinfo{person}{Kay
  K{\"{u}}hne}, {and} \bibinfo{person}{J{\"{u}}rgen Beyerer}.}
  \bibinfo{year}{2016}\natexlab{}.
\newblock \showarticletitle{PrivacyInsight: The Next Generation Privacy
  Dashboard}. In \bibinfo{booktitle}{\emph{Privacy Technologies and Policy}},
  \bibfield{editor}{\bibinfo{person}{Stefan Schiffner},
  \bibinfo{person}{Jetzabel Serna}, \bibinfo{person}{Demosthenes Ikonomou},
  {and} \bibinfo{person}{Kai Rannenberg}} (Eds.). \bibinfo{publisher}{Springer
  International Publishing}, \bibinfo{address}{Cham},
  \bibinfo{pages}{135--152}.
\newblock
\showISBNx{978-3-319-44760-5}


\bibitem[\protect\citeauthoryear{Chang, Li, Zhang, Du, Cao, and Zhu}{Chang
  et~al\mbox{.}}{2019}]%
        {chang-policy-extraction}
\bibfield{author}{\bibinfo{person}{Cheng Chang}, \bibinfo{person}{Huaxin Li},
  \bibinfo{person}{Yichi Zhang}, \bibinfo{person}{Suguo Du},
  \bibinfo{person}{Hui Cao}, {and} \bibinfo{person}{Haojin Zhu}.}
  \bibinfo{year}{2019}\natexlab{}.
\newblock \showarticletitle{Automated and Personalized Privacy Policy
  Extraction Under GDPR Consideration}. In \bibinfo{booktitle}{\emph{Wireless
  Algorithms, Systems, and Applications}},
  \bibfield{editor}{\bibinfo{person}{Edoardo~S. Biagioni}, \bibinfo{person}{Yao
  Zheng}, {and} \bibinfo{person}{Siyao Cheng}} (Eds.).
  \bibinfo{publisher}{Springer International Publishing},
  \bibinfo{address}{Cham}, \bibinfo{pages}{43--54}.
\newblock
\showISBNx{978-3-030-23597-0}


\bibitem[\protect\citeauthoryear{Costante, den Hartog, and
  Petkovi{\'{c}}}{Costante et~al\mbox{.}}{2013}]%
        {costante2013extraction}
\bibfield{author}{\bibinfo{person}{Elisa Costante}, \bibinfo{person}{Jerry den
  Hartog}, {and} \bibinfo{person}{Milan Petkovi{\'{c}}}.}
  \bibinfo{year}{2013}\natexlab{}.
\newblock \showarticletitle{What Websites Know About You}. In
  \bibinfo{booktitle}{\emph{Data Privacy Management and Autonomous Spontaneous
  Security}}, \bibfield{editor}{\bibinfo{person}{Roberto Di~Pietro},
  \bibinfo{person}{Javier Herranz}, \bibinfo{person}{Ernesto Damiani}, {and}
  \bibinfo{person}{Radu State}} (Eds.). \bibinfo{publisher}{Springer Berlin
  Heidelberg}, \bibinfo{address}{Berlin, Heidelberg},
  \bibinfo{pages}{146--159}.
\newblock
\showISBNx{978-3-642-35890-6}


\bibitem[\protect\citeauthoryear{Cranor}{Cranor}{2003}]%
        {cranor2003}
\bibfield{author}{\bibinfo{person}{Lorrie~Faith Cranor}.}
  \bibinfo{year}{2003}\natexlab{}.
\newblock \showarticletitle{P3P: Making privacy policies more useful}.
\newblock \bibinfo{journal}{\emph{IEEE Security \& Privacy}}
  \bibinfo{volume}{1} (\bibinfo{year}{2003}), \bibinfo{pages}{50--55}.
\newblock
Issue 6.
\urldef\tempurl%
\url{https://doi.org/10.1109/MSECP.2003.1253568}
\showDOI{\tempurl}


\bibitem[\protect\citeauthoryear{{European Parliament \& Council}}{{European
  Parliament \& Council}}{2016}]%
        {gdpr}
\bibfield{author}{\bibinfo{person}{{European Parliament \& Council}}.}
  \bibinfo{year}{2016}\natexlab{}.
\newblock \bibinfo{title}{Regulation ({{EU}}) 2016/679 of the {{European
  Parliament}} and of the {{Council}} of 27 {{April}} 2016 on the Protection of
  Natural Persons with Regard to the Processing of Personal Data and on the
  Free Movement of Such Data, and Repealing {{Directive}} 95/46/{{EC}}
  ({{General Data Protection Regulation}})}.
\newblock
\newblock


\bibitem[\protect\citeauthoryear{Fukushima, Nakamura, Ikeda, and
  Kiyomoto}{Fukushima et~al\mbox{.}}{2018}]%
        {fukushima2018icon-representation}
\bibfield{author}{\bibinfo{person}{Keishiro Fukushima}, \bibinfo{person}{Toru
  Nakamura}, \bibinfo{person}{Daisuke Ikeda}, {and} \bibinfo{person}{Shinsaku
  Kiyomoto}.} \bibinfo{year}{2018}\natexlab{}.
\newblock \showarticletitle{Challenges in Classifying Privacy Policies by
  Machine Learning with Word-Based Features}. In
  \bibinfo{booktitle}{\emph{Proceedings of the 2nd International Conference on
  Cryptography, Security and Privacy}} \emph{(\bibinfo{series}{ICCSP 2018})}.
  \bibinfo{publisher}{Association for Computing Machinery},
  \bibinfo{address}{New York, NY, USA}, \bibinfo{pages}{62–66}.
\newblock
\showISBNx{9781450363617}
\urldef\tempurl%
\url{https://doi.org/10.1145/3199478.3199486}
\showDOI{\tempurl}


\bibitem[\protect\citeauthoryear{Gerl and Meier}{Gerl and Meier}{2019}]%
        {gerl2019}
\bibfield{author}{\bibinfo{person}{Armin Gerl} {and} \bibinfo{person}{Bianca
  Meier}.} \bibinfo{year}{2019}\natexlab{}.
\newblock \showarticletitle{The Layered Privacy Language Art. 12--14 GDPR
  Extension--Privacy Enhancing User Interfaces}.
\newblock \bibinfo{journal}{\emph{Datenschutz und Datensicherheit-DuD}}
  \bibinfo{volume}{43}, \bibinfo{number}{12} (\bibinfo{year}{2019}),
  \bibinfo{pages}{747--752}.
\newblock


\bibitem[\protect\citeauthoryear{Harkous, Fawaz, Lebret, Schaub, Shin, and
  Aberer}{Harkous et~al\mbox{.}}{2018}]%
        {polisis}
\bibfield{author}{\bibinfo{person}{Hamza Harkous}, \bibinfo{person}{Kassem
  Fawaz}, \bibinfo{person}{R{\'e}mi Lebret}, \bibinfo{person}{Florian Schaub},
  \bibinfo{person}{Kang~G. Shin}, {and} \bibinfo{person}{Karl Aberer}.}
  \bibinfo{year}{2018}\natexlab{}.
\newblock \showarticletitle{Polisis: Automated Analysis and Presentation of
  Privacy Policies Using Deep Learning}. In \bibinfo{booktitle}{\emph{27th
  {USENIX} Security Symposium ({USENIX} Security 18)}}.
  \bibinfo{publisher}{{USENIX} Association}, \bibinfo{address}{Baltimore, MD},
  \bibinfo{pages}{531--548}.
\newblock
\showISBNx{978-1-939133-04-5}
\urldef\tempurl%
\url{https://www.usenix.org/conference/usenixsecurity18/presentation/harkous}
\showURL{%
\tempurl}


\bibitem[\protect\citeauthoryear{Harkous, Fawaz, Shin, and Aberer}{Harkous
  et~al\mbox{.}}{2016}]%
        {pribots}
\bibfield{author}{\bibinfo{person}{Hamza Harkous}, \bibinfo{person}{Kassem
  Fawaz}, \bibinfo{person}{Kang~G. Shin}, {and} \bibinfo{person}{Karl Aberer}.}
  \bibinfo{year}{2016}\natexlab{}.
\newblock \showarticletitle{PriBots: Conversational Privacy with Chatbots}. In
  \bibinfo{booktitle}{\emph{Twelfth Symposium on Usable Privacy and Security
  ({SOUPS} 2016)}}. \bibinfo{publisher}{{USENIX} Association},
  \bibinfo{address}{Denver, CO}, 6.
\newblock
\urldef\tempurl%
\url{https://www.usenix.org/conference/soups2016/workshop-program/wfpn/presentation/harkous}
\showURL{%
\tempurl}


\bibitem[\protect\citeauthoryear{Hedbom}{Hedbom}{2009}]%
        {hedbom2009}
\bibfield{author}{\bibinfo{person}{Hans Hedbom}.}
  \bibinfo{year}{2009}\natexlab{}.
\newblock \showarticletitle{A Survey on Transparency Tools for Enhancing
  Privacy}. In \bibinfo{booktitle}{\emph{The Future of Identity in the
  Information Society}}, \bibfield{editor}{\bibinfo{person}{Vashek
  Maty{\'a}{\v{s}}}, \bibinfo{person}{Simonen Fischer-H{\"u}bner},
  \bibinfo{person}{Daniel Cvr{\v{c}}ek}, {and} \bibinfo{person}{Petr
  {\v{S}}venda}} (Eds.). \bibinfo{publisher}{Springer Berlin Heidelberg},
  \bibinfo{address}{Berlin, Heidelberg}, \bibinfo{pages}{67--82}.
\newblock
\showISBNx{978-3-642-03315-5}


\bibitem[\protect\citeauthoryear{Holtz, Nocun, and Hansen}{Holtz
  et~al\mbox{.}}{2011a}]%
        {holtz}
\bibfield{author}{\bibinfo{person}{Leif-Erik Holtz}, \bibinfo{person}{Katharina
  Nocun}, {and} \bibinfo{person}{Marit Hansen}.}
  \bibinfo{year}{2011}\natexlab{a}.
\newblock \showarticletitle{Towards Displaying Privacy Information with Icons}.
  In \bibinfo{booktitle}{\emph{Privacy and Identity Management for Life}},
  \bibfield{editor}{\bibinfo{person}{Simone Fischer-H{\"u}bner},
  \bibinfo{person}{Penny Duquenoy}, \bibinfo{person}{Marit Hansen},
  \bibinfo{person}{Ronald Leenes}, {and} \bibinfo{person}{Ge~Zhang}} (Eds.).
  \bibinfo{publisher}{Springer Berlin Heidelberg}, \bibinfo{address}{Berlin,
  Heidelberg}, \bibinfo{pages}{338--348}.
\newblock
\showISBNx{978-3-642-20769-3}


\bibitem[\protect\citeauthoryear{Holtz, Zwingelberg, and Hansen}{Holtz
  et~al\mbox{.}}{2011b}]%
        {Holtz2011}
\bibfield{author}{\bibinfo{person}{Leif-Erik Holtz}, \bibinfo{person}{Harald
  Zwingelberg}, {and} \bibinfo{person}{Marit Hansen}.}
  \bibinfo{year}{2011}\natexlab{b}.
\newblock \bibinfo{booktitle}{\emph{Privacy Policy Icons}}.
\newblock \bibinfo{publisher}{Springer Berlin Heidelberg},
  \bibinfo{address}{Berlin, Heidelberg}. 279--285 pages.
\newblock
\showISBNx{978-3-642-20317-6}
\urldef\tempurl%
\url{https://doi.org/10.1007/978-3-642-20317-6_15}
\showDOI{\tempurl}


\bibitem[\protect\citeauthoryear{(IAB)}{(IAB)}{2020}]%
        {iab-tcf}
\bibfield{author}{\bibinfo{person}{Interactive Advertising~Bureau (IAB)}.}
  \bibinfo{year}{2020}\natexlab{}.
\newblock \bibinfo{title}{TCF – Transparency \& Consent Framework}.
\newblock
\newblock
\urldef\tempurl%
\url{https://iabeurope.eu/transparency-consent-framework/}
\showURL{%
\tempurl}


\bibitem[\protect\citeauthoryear{ISO IEC 14977 996 E}{ISO IEC 14977 996
  E}{1996}]%
        {standard1996ebnf}
ISO IEC 14977 996 E \bibinfo{year}{1996}\natexlab{}.
\newblock \bibinfo{booktitle}{\emph{EBNF Syntax Specification}}.
\newblock \bibinfo{type}{Standard}. \bibinfo{institution}{International
  Organization for Standardization}, \bibinfo{address}{Geneva, CH}.
\newblock


\bibitem[\protect\citeauthoryear{Janic, Wijbenga, and Veugen}{Janic
  et~al\mbox{.}}{2013}]%
        {janic2013}
\bibfield{author}{\bibinfo{person}{Milena Janic}, \bibinfo{person}{Jan~Pieter
  Wijbenga}, {and} \bibinfo{person}{Thijs Veugen}.}
  \bibinfo{year}{2013}\natexlab{}.
\newblock \showarticletitle{Transparency enhancing tools (TETs): An overview}.
  In \bibinfo{booktitle}{\emph{Proceedings of the Workshop on Socio-Technical
  Aspects in Security and Trust, STAST}}. \bibinfo{publisher}{IEEE},
  \bibinfo{address}{New Orleans, LA, USA}, \bibinfo{pages}{18--25}.
\newblock
\showISBNx{9780769550657}
\showISSN{23251689}
\urldef\tempurl%
\url{https://doi.org/10.1109/STAST.2013.11}
\showDOI{\tempurl}


\bibitem[\protect\citeauthoryear{Kakavas}{Kakavas}{2016}]%
        {creepy}
\bibfield{author}{\bibinfo{person}{Ioannis Kakavas}.}
  \bibinfo{year}{2016}\natexlab{}.
\newblock \bibinfo{title}{Creepy. A geolocation OSINT tool.}
\newblock
\newblock
\urldef\tempurl%
\url{https://www.geocreepy.com/}
\showURL{%
\tempurl}


\bibitem[\protect\citeauthoryear{Kaminski}{Kaminski}{2020}]%
        {kaminski2020recent}
\bibfield{author}{\bibinfo{person}{Margot Kaminski}.}
  \bibinfo{year}{2020}\natexlab{}.
\newblock \showarticletitle{Law and Technology. A recent renaissance in privacy
  law}.
\newblock \bibinfo{journal}{\emph{Commun. ACM}} \bibinfo{volume}{63},
  \bibinfo{number}{9} (\bibinfo{year}{2020}), \bibinfo{pages}{24--27}.
\newblock
\urldef\tempurl%
\url{https://doi.org/10.1145/3411049}
\showDOI{\tempurl}


\bibitem[\protect\citeauthoryear{Karegar, Pulls, and
  Fischer-H{\"u}bner}{Karegar et~al\mbox{.}}{2016}]%
        {Karegar2016}
\bibfield{author}{\bibinfo{person}{Farzaneh Karegar}, \bibinfo{person}{Tobias
  Pulls}, {and} \bibinfo{person}{Simone Fischer-H{\"u}bner}.}
  \bibinfo{year}{2016}\natexlab{}.
\newblock \bibinfo{booktitle}{\emph{Visualizing Exports of Personal Data by
  Exercising the Right of Data Portability in the Data Track - Are People Ready
  for This?}}
\newblock \bibinfo{publisher}{Springer International Publishing},
  \bibinfo{address}{Cham}, \bibinfo{pages}{164--181}.
\newblock
\showISBNx{978-3-319-55783-0}
\urldef\tempurl%
\url{https://doi.org/10.1007/978-3-319-55783-0_12}
\showDOI{\tempurl}


\bibitem[\protect\citeauthoryear{Kirrane, Fernández, Bonatti, Petrova, Sauro,
  and Schlehahn}{Kirrane et~al\mbox{.}}{2019}]%
        {special-usage-language}
\bibfield{author}{\bibinfo{person}{Sabrina Kirrane}, \bibinfo{person}{Javier~D.
  Fernández}, \bibinfo{person}{Piero Bonatti}, \bibinfo{person}{Iliana~Mineva
  Petrova}, \bibinfo{person}{Luigi Sauro}, {and} \bibinfo{person}{Eva
  Schlehahn}.} \bibinfo{year}{2019}\natexlab{}.
\newblock \bibinfo{title}{The SPECIAL Usage Policy Language}.
\newblock
\newblock
\urldef\tempurl%
\url{https://ai.wu.ac.at/policies/policylanguage/}
\showURL{%
\tempurl}


\bibitem[\protect\citeauthoryear{Knoke and Yang}{Knoke and Yang}{2019}]%
        {knoke2019social}
\bibfield{author}{\bibinfo{person}{David Knoke} {and} \bibinfo{person}{Song
  Yang}.} \bibinfo{year}{2019}\natexlab{}.
\newblock \bibinfo{booktitle}{\emph{Social network analysis}}.
  Vol.~\bibinfo{volume}{154}.
\newblock \bibinfo{publisher}{Sage Publications}.
\newblock


\bibitem[\protect\citeauthoryear{Linden, Khandelwal, Harkous, and Fawaz}{Linden
  et~al\mbox{.}}{2018}]%
        {Linden2018}
\bibfield{author}{\bibinfo{person}{Thomas Linden}, \bibinfo{person}{Rishabh
  Khandelwal}, \bibinfo{person}{Hamza Harkous}, {and} \bibinfo{person}{Kassem
  Fawaz}.} \bibinfo{year}{2018}\natexlab{}.
\newblock \bibinfo{title}{The Privacy Policy Landscape After the GDPR}.
\newblock
\newblock
\urldef\tempurl%
\url{http://arxiv.org/abs/1809.08396}
\showURL{%
\tempurl}


\bibitem[\protect\citeauthoryear{McDonald and Cranor}{McDonald and
  Cranor}{2008}]%
        {donald-cranor2008cost-reading}
\bibfield{author}{\bibinfo{person}{Aleecia~M McDonald} {and}
  \bibinfo{person}{Lorrie~Faith Cranor}.} \bibinfo{year}{2008}\natexlab{}.
\newblock \showarticletitle{The Cost of Reading Privacy Policies}.
\newblock \bibinfo{journal}{\emph{Journal of Law and Policy for the Information
  Society}}  \bibinfo{volume}{4} (\bibinfo{year}{2008}),
  \bibinfo{pages}{543--568}.
\newblock


\bibitem[\protect\citeauthoryear{Mernik, Heering, and Sloane}{Mernik
  et~al\mbox{.}}{2005}]%
        {whenAndHow2005}
\bibfield{author}{\bibinfo{person}{Marjan Mernik}, \bibinfo{person}{Jan
  Heering}, {and} \bibinfo{person}{Anthony~M. Sloane}.}
  \bibinfo{year}{2005}\natexlab{}.
\newblock \showarticletitle{When and How to Develop Domain-Specific Languages}.
\newblock \bibinfo{journal}{\emph{ACM Comput. Surv.}} \bibinfo{volume}{37},
  \bibinfo{number}{4} (\bibinfo{date}{Dec.} \bibinfo{year}{2005}),
  \bibinfo{pages}{316–344}.
\newblock
\showISSN{0360-0300}
\urldef\tempurl%
\url{https://doi.org/10.1145/1118890.1118892}
\showDOI{\tempurl}


\bibitem[\protect\citeauthoryear{Obar and Oeldorf-Hirsch}{Obar and
  Oeldorf-Hirsch}{2018}]%
        {obar2018}
\bibfield{author}{\bibinfo{person}{Jonathan~A. Obar} {and}
  \bibinfo{person}{Anne Oeldorf-Hirsch}.} \bibinfo{year}{2018}\natexlab{}.
\newblock \showarticletitle{The biggest lie on the Internet: ignoring the
  privacy policies and terms of service policies of social networking
  services}.
\newblock \bibinfo{journal}{\emph{Information, Communication \& Society}}
  \bibinfo{volume}{23}, \bibinfo{number}{1} (\bibinfo{year}{2018}),
  \bibinfo{pages}{128--147}.
\newblock
\urldef\tempurl%
\url{https://doi.org/10.1080/1369118X.2018.1486870}
\showDOI{\tempurl}


\bibitem[\protect\citeauthoryear{Paige, Ostroff, and Brooke}{Paige
  et~al\mbox{.}}{2000}]%
        {paige2000principles}
\bibfield{author}{\bibinfo{person}{Richard~F. Paige},
  \bibinfo{person}{Jonathan~S. Ostroff}, {and} \bibinfo{person}{Phillip~J
  Brooke}.} \bibinfo{year}{2000}\natexlab{}.
\newblock \showarticletitle{Principles for modeling language design}.
\newblock \bibinfo{journal}{\emph{Information and Software Technology}}
  \bibinfo{volume}{42}, \bibinfo{number}{10} (\bibinfo{year}{2000}),
  \bibinfo{pages}{665--675}.
\newblock


\bibitem[\protect\citeauthoryear{Pallas, Ulbricht, Tai, Peikert, Reppenhagen,
  Wenzel, Wille, and Wolf}{Pallas et~al\mbox{.}}{2020}]%
        {pbac2020}
\bibfield{author}{\bibinfo{person}{Frank Pallas}, \bibinfo{person}{Max-R.
  Ulbricht}, \bibinfo{person}{Stefan Tai}, \bibinfo{person}{Thomas Peikert},
  \bibinfo{person}{Marcel Reppenhagen}, \bibinfo{person}{Daniel Wenzel},
  \bibinfo{person}{Paul Wille}, {and} \bibinfo{person}{Karl Wolf}.}
  \bibinfo{year}{2020}\natexlab{}.
\newblock \showarticletitle{Towards Application-Layer Purpose-Based Access
  Control}. In \bibinfo{booktitle}{\emph{Proceedings of the 35th Annual ACM
  Symposium on Applied Computing}} \emph{(\bibinfo{series}{SAC '20})}.
  \bibinfo{publisher}{Association for Computing Machinery},
  \bibinfo{address}{New York, NY, USA}, \bibinfo{pages}{1288–1296}.
\newblock
\showISBNx{9781450368667}
\urldef\tempurl%
\url{https://doi.org/10.1145/3341105.3375764}
\showDOI{\tempurl}


\bibitem[\protect\citeauthoryear{Pezoa, Reutter, Suarez, Ugarte, and
  Vrgoc}{Pezoa et~al\mbox{.}}{2016}]%
        {jsonschema2016}
\bibfield{author}{\bibinfo{person}{Felipe Pezoa}, \bibinfo{person}{Juan~L.
  Reutter}, \bibinfo{person}{Fernando Suarez}, \bibinfo{person}{Martin Ugarte},
  {and} \bibinfo{person}{Domagoj Vrgoc}.} \bibinfo{year}{2016}\natexlab{}.
\newblock \showarticletitle{Foundations of JSON Schema}. In
  \bibinfo{booktitle}{\emph{Proceedings of the 25th International Conference on
  World Wide Web}} \emph{(\bibinfo{series}{WWW '16})}.
  \bibinfo{publisher}{International World Wide Web Conferences Steering
  Committee}, \bibinfo{address}{Republic and Canton of Geneva, CHE},
  \bibinfo{pages}{263–273}.
\newblock
\showISBNx{9781450341431}
\urldef\tempurl%
\url{https://doi.org/10.1145/2872427.2883029}
\showDOI{\tempurl}


\bibitem[\protect\citeauthoryear{Princiya}{Princiya}{2019}]%
        {lightbeam}
\bibfield{author}{\bibinfo{person}{Princiya}.} \bibinfo{year}{2019}\natexlab{}.
\newblock \bibinfo{title}{Mozilla Firefox Lightbeam}.
\newblock
\newblock
\urldef\tempurl%
\url{https://addons.mozilla.org/firefox/addon/lightbeam-3-0/}
\showURL{%
\tempurl}


\bibitem[\protect\citeauthoryear{Raschke, K{\"u}pper, Drozd, and
  Kirrane}{Raschke et~al\mbox{.}}{2018}]%
        {Raschke2018}
\bibfield{author}{\bibinfo{person}{Philip Raschke}, \bibinfo{person}{Axel
  K{\"u}pper}, \bibinfo{person}{Olha Drozd}, {and} \bibinfo{person}{Sabrina
  Kirrane}.} \bibinfo{year}{2018}\natexlab{}.
\newblock \showarticletitle{Designing a GDPR-Compliant and Usable Privacy
  Dashboard}.
\newblock In \bibinfo{booktitle}{\emph{Privacy and Identity Management. The
  Smart Revolution}}, \bibfield{editor}{\bibinfo{person}{Marit Hansen},
  \bibinfo{person}{Eleni Kosta}, \bibinfo{person}{Igor Nai-Fovino}, {and}
  \bibinfo{person}{Simone Fischer-H{\"u}bner}} (Eds.).
  \bibinfo{publisher}{Springer International Publishing},
  \bibinfo{address}{Cham}, \bibinfo{pages}{221--236}.
\newblock
\showISBNx{978-3-319-92925-5}
\urldef\tempurl%
\url{https://doi.org/10.1007/978-3-319-92925-5_14}
\showDOI{\tempurl}


\bibitem[\protect\citeauthoryear{Reidenberg, Breaux, Cranor, French, Grannis,
  Graves, Liu, McDonald, Norton, and Ramanath}{Reidenberg
  et~al\mbox{.}}{2015}]%
        {ReidenbergDisagreeablePrivacyPolicies2015}
\bibfield{author}{\bibinfo{person}{Joel~R. Reidenberg}, \bibinfo{person}{Travis
  Breaux}, \bibinfo{person}{Lorrie~Faith Cranor}, \bibinfo{person}{Brian
  French}, \bibinfo{person}{Amanda Grannis}, \bibinfo{person}{James~T. Graves},
  \bibinfo{person}{Fei Liu}, \bibinfo{person}{Aleecia McDonald},
  \bibinfo{person}{Thomas~B. Norton}, {and} \bibinfo{person}{Rohan Ramanath}.}
  \bibinfo{year}{2015}\natexlab{}.
\newblock \showarticletitle{Disagreeable {{Privacy Policies}}: {{Mismatches}}
  between {{Meaning}} and {{Users}}' {{Understanding}}}.
\newblock \bibinfo{journal}{\emph{Berkeley Technology Law Journal}}
  \bibinfo{volume}{30} (\bibinfo{year}{2015}), \bibinfo{pages}{39}.
\newblock


\bibitem[\protect\citeauthoryear{{Rossi} and {Palmirani}}{{Rossi} and
  {Palmirani}}{2020}]%
        {rossi}
\bibfield{author}{\bibinfo{person}{Arianna {Rossi}} {and}
  \bibinfo{person}{Monica {Palmirani}}.} \bibinfo{year}{2020}\natexlab{}.
\newblock \showarticletitle{Can Visual Design Provide Legal Transparency? The
  Challenges for Successful Implementation of Icons for Data Protection}.
\newblock \bibinfo{journal}{\emph{Design Issues}} \bibinfo{volume}{36},
  \bibinfo{number}{3} (\bibinfo{year}{2020}), \bibinfo{pages}{82--96}.
\newblock


\bibitem[\protect\citeauthoryear{Sadeh, Acquisti, Breaux, Cranor, Mcdonalda,
  Reidenbergb, Smith, Liu, Russellb, Schaub, and Wilson}{Sadeh
  et~al\mbox{.}}{2013}]%
        {u3p}
\bibfield{author}{\bibinfo{person}{Norman Sadeh}, \bibinfo{person}{Ro
  Acquisti}, \bibinfo{person}{Travis~D. Breaux}, \bibinfo{person}{Lorrie~Faith
  Cranor}, \bibinfo{person}{Aleecia~M. Mcdonalda}, \bibinfo{person}{Joel~R.
  Reidenbergb}, \bibinfo{person}{Noah~A. Smith}, \bibinfo{person}{Fei Liu},
  \bibinfo{person}{N.~Cameron Russellb}, \bibinfo{person}{Florian Schaub},
  {and} \bibinfo{person}{Shomir Wilson}.} \bibinfo{year}{2013}\natexlab{}.
\newblock \bibinfo{title}{The Usable Privacy Policy Project: Combining
  Crowdsourcing, Machine Learning and Natural Language Processing to
  Semi-Automatically Answer Those Privacy Questions Users Care About}.
\newblock
\newblock
\urldef\tempurl%
\url{http://ra.adm.cs.cmu.edu/anon/usr0/ftp/home/anon/isr2013/CMU-ISR-13-119.pdf}
\showURL{%
\tempurl}


\bibitem[\protect\citeauthoryear{{The State of California}}{{The State of
  California}}{2018}]%
        {ccpa}
\bibfield{author}{\bibinfo{person}{{The State of California}}.}
  \bibinfo{year}{2018}\natexlab{}.
\newblock \bibinfo{title}{California Consumer Privacy Act of 2018}.
\newblock
\newblock
\urldef\tempurl%
\url{http://leginfo.legislature.ca.gov/faces/codes_displayText.xhtml?division=3.&part=4.&lawCode=CIV&title=1.81.5}
\showURL{%
\tempurl}


\bibitem[\protect\citeauthoryear{Trepte, Teutsch, Masur, Eicher, Fischer,
  Hennh{\"o}fer, and Lind}{Trepte et~al\mbox{.}}{2015}]%
        {trepte2015}
\bibfield{author}{\bibinfo{person}{Sabine Trepte}, \bibinfo{person}{Doris
  Teutsch}, \bibinfo{person}{Philipp~K. Masur}, \bibinfo{person}{Carolin
  Eicher}, \bibinfo{person}{Mona Fischer}, \bibinfo{person}{Alisa
  Hennh{\"o}fer}, {and} \bibinfo{person}{Fabienne Lind}.}
  \bibinfo{year}{2015}\natexlab{}.
\newblock \bibinfo{booktitle}{\emph{Do People Know About Privacy and Data
  Protection Strategies? Towards the “Online Privacy Literacy Scale”
  (OPLIS)}}.
\newblock \bibinfo{publisher}{Springer Netherlands},
  \bibinfo{address}{Dordrecht}, \bibinfo{pages}{333--365}.
\newblock
\showISBNx{978-94-017-9385-8}
\urldef\tempurl%
\url{https://doi.org/10.1007/978-94-017-9385-8}
\showDOI{\tempurl}


\bibitem[\protect\citeauthoryear{Ulbricht and Pallas}{Ulbricht and
  Pallas}{2018}]%
        {UlbrichtYaPPLLightweightPrivacy2018}
\bibfield{author}{\bibinfo{person}{Max.-R. Ulbricht} {and}
  \bibinfo{person}{Frank Pallas}.} \bibinfo{year}{2018}\natexlab{}.
\newblock \showarticletitle{{{YaPPL}} - {{A Lightweight Privacy Preference
  Language}} for {{Legally Sufficient}} and {{Automated Consent Provision}} in
  {{IoT Scenarios}}}. In \bibinfo{booktitle}{\emph{Proceedings of {{Data
  Privacy Management}} 2018}} \emph{(\bibinfo{series}{LNCS})},
  \bibfield{editor}{\bibinfo{person}{{Giovanni Livraga}} {and}
  \bibinfo{person}{{Ruben Rios}}} (Eds.), Vol.~\bibinfo{volume}{11025}.
  \bibinfo{publisher}{{Springer International Publishing}},
  \bibinfo{pages}{329--344}.
\newblock


\bibitem[\protect\citeauthoryear{Warcha{\l}}{Warcha{\l}}{2012}]%
        {warchal2012using}
\bibfield{author}{\bibinfo{person}{{\L}ukasz Warcha{\l}}.}
  \bibinfo{year}{2012}\natexlab{}.
\newblock \showarticletitle{Using Neo4j graph database in social network
  analysis}.
\newblock \bibinfo{journal}{\emph{Studia Informatica}} \bibinfo{volume}{33},
  \bibinfo{number}{2A} (\bibinfo{year}{2012}), \bibinfo{pages}{271--279}.
\newblock


\bibitem[\protect\citeauthoryear{Wetherbee, Nardone, Rathod, and
  Kodali}{Wetherbee et~al\mbox{.}}{2018}]%
        {Wetherbee2018}
\bibfield{author}{\bibinfo{person}{Jonathan Wetherbee},
  \bibinfo{person}{Massimo Nardone}, \bibinfo{person}{Chirag Rathod}, {and}
  \bibinfo{person}{Raghu Kodali}.} \bibinfo{year}{2018}\natexlab{}.
\newblock \bibinfo{booktitle}{\emph{EJB, Web Services, and Microservices}}.
\newblock \bibinfo{publisher}{Apress}, \bibinfo{address}{Berkeley, CA},
  \bibinfo{pages}{265--317}.
\newblock
\showISBNx{978-1-4842-3573-7}
\urldef\tempurl%
\url{https://doi.org/10.1007/978-1-4842-3573-7_6}
\showDOI{\tempurl}


\bibitem[\protect\citeauthoryear{Xu and Zhu}{Xu and Zhu}{2015}]%
        {semadroid}
\bibfield{author}{\bibinfo{person}{Zhi Xu} {and} \bibinfo{person}{Sencun Zhu}.}
  \bibinfo{year}{2015}\natexlab{}.
\newblock \showarticletitle{SemaDroid: A Privacy-Aware Sensor Management
  Framework for Smartphones}. In \bibinfo{booktitle}{\emph{Proceedings of the
  5th ACM Conference on Data and Application Security and Privacy}}
  \emph{(\bibinfo{series}{CODASPY '15})}. \bibinfo{publisher}{Association for
  Computing Machinery}, \bibinfo{address}{New York, NY, USA},
  \bibinfo{pages}{61–72}.
\newblock
\showISBNx{9781450331913}
\urldef\tempurl%
\url{https://doi.org/10.1145/2699026.2699114}
\showDOI{\tempurl}


\bibitem[\protect\citeauthoryear{Zimmermann}{Zimmermann}{2015}]%
        {zimmermann2015}
\bibfield{author}{\bibinfo{person}{Christian Zimmermann}.}
  \bibinfo{year}{2015}\natexlab{}.
\newblock \bibinfo{title}{A categorization of transparency-enhancing
  technologies}.
\newblock \bibinfo{howpublished}{arXiv}.
\newblock
\urldef\tempurl%
\url{https://arxiv.org/abs/1507.04914}
\showURL{%
\tempurl}


\end{thebibliography}

\end{document}